# Spatial Wavefunctions of Spin

T. Peter Rakitzis

*Department of Physics, University of Crete, 70013 Heraklion-Crete, Greece*

*IESL-FORTH, N. Plastira 100, Heraklion-Crete 71110, Greece*

*Email:* ptr@iesl.forth.gr

11/1/2025

We present an alternative formulation of quantum mechanical angular momentum, based on spatial wavefunctions that depend on the Euler angles $\varphi, \theta, \chi$, and have an additional internal projection $n$. The wavefunctions are Wigner D-functions, $D^s_{n\,m}(\varphi, \theta, \chi)$, for which the body-fixed projection quantum number $n$ has the unusual value $n = |\mathbf{s}| = \sqrt{s(s+1)} = \mathcal{s}$, or $n = 0$. We show that the states $D^s_{\mathcal{s}\,m}(\varphi, \theta, \chi)$ of elementary particles with spin $s$ give a gyromagnetic ratio of $g = 2$ for $s > 0$, and we identify these as the spatial angular-momentum wavefunctions of known fundamental charged particles with spin. All known Standard-Model particles can be categorized with either value $n = \mathcal{s}$ or $n = 0$, and all known particle reactions are consistent with the conservation of its projection in the internal frame, and with internal-frame Clebsch-Gordan coefficients of unity. Therefore, we make the case that the $D^s_{n\,m}(\varphi, \theta, \chi)$ are useful as spatial wavefunctions for spin. Some implications and new predictions related to the quantum number $n$ for fundamental particles are discussed, such as the proposed Dirac-fermion nature of the neutrino, the explanation of some Standard-Model structure, and some proposed dark-matter candidates.

# I. INTRODUCTION

In conventional quantum-mechanical angular-momentum theory, there are spatial wavefunctions for describing the angular position of an orbiting particle in a central field, the spherical harmonics $Y_m^l(\theta, \varphi)$ (where $l$ is the angular momentum quantum number, and $m$ is the azimuthal quantum number, the projection of the angular momentum $\boldsymbol{l}$ along the space-fixed Z axis). The $Y_m^l(\theta, \varphi)$ give the probability amplitude of finding the particle at the angular location described by the angles $\theta$ and $\varphi$. However, no analogous spatial wavefunction exists for angular momentum or spin in conventional quantum mechanics, which would describe the angular distribution of the angular momentum in terms of spatial coordinates. In fact, it is generally believed that spatial wavefunctions for spin do not exist. For example, in a well-known textbook [1] it is stated that: "… the electron also carries another form of angular momentum, which has nothing to do with motion in space (and which is not, therefore, described by any function of the position variables $r, \theta, \varphi$) but which is somewhat analogous to classical spin (and for which, therefore, we use the same word)".

The aim of this paper is to show that spatial wavefunctions of spin exist and are useful, in the form of a Wigner D-function, $D_{n\,m}^s(\varphi, \theta, \chi)$, for which the body-fixed projection quantum number $n$ has the unusual value $n = |\boldsymbol{s}| = \sqrt{s(s+1)} = \boldsymbol{s}$, or $n = 0$ (even for $s = 1/2$). We first give a practical review of the usual Wigner $D_{m'\,m}^j(\varphi, \theta, \chi)$ functions (for which $m'$ must range from $-j$ to $+j$ in integer steps), and then give some arguments for extending the value to $m' = \sqrt{j(j+1)}$, for angular momentum spatial wavefunctions, which describe the spatial distribution of the angular momentum vector $\boldsymbol{j}$.

For the description of the rotation of bodies with internal structure (such as molecules), the Wigner $D_{m'\,m}^j(\varphi, \theta, \chi)$ functions are spatial wavefunctions that describe the angular distribution of the body-fixed $z'$ axis (which is parallel to the principle axis of the molecule) for given projections $m$ and $m'$ of $j$ along Z and $z'$, respectively, and $\varphi, \theta, \chi$ are the Euler angles. For the $D_{m'\,m}^j(\varphi, \theta, \chi)$ to be non-divergent and normalizable $m$ and $m'$ must range from $-j$ to $+j$ in integer steps. We give three examples of using $D_{m'\,m}^j(\varphi, \theta, \chi)$ as spatial wavefunctions, culminating in a proposal for a spatial wavefunction for the spin vector $\boldsymbol{s}$.

First, we consider a symmetric-top molecule

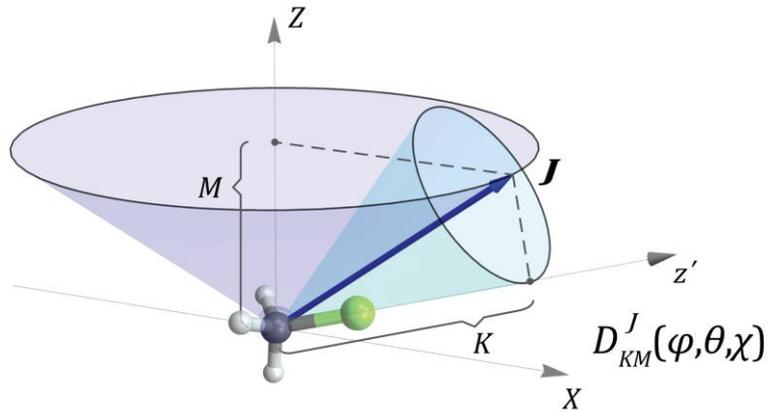

*Figure 1:* Vector-Model representation of the symmetric top molecule CH₃Cl in the $|JKM\rangle$ state. The angular momentum J projects M along the space-fixed Z axis, and K along the body-fixed z axis (which is parallel to the principle axis of the molecule).

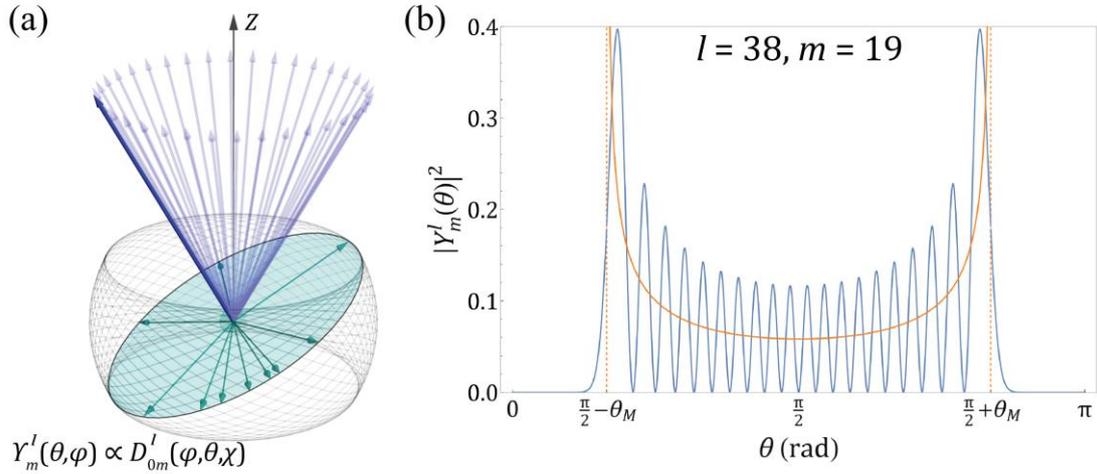

**Figure 2:** *(a) Vector-Model description of the $Y_m^l(\theta, \varphi)$ state: the orbital angular momentum vector **l** projects m along Z, and is delocalized around Z in a cone. The orbital angular momentum **l** also projects 0 along z' (which is parallel to the position of the particle **r**) and is perpendicular to the orbital plane of the orbiting particle (which precesses along with **l**). (b) Exact values of $|Y_m^l(\theta, \varphi)|^2$ (in blue) are compared to the classical VM prediction of the probability distribution of the particle; the $Y_m^l(\theta, \varphi)$ oscillate within the classically-allowed region, and decay exponentially outside it.*

(such as CH$_3$Cl) in the state $|JKM\rangle \equiv D_{KM}^J(\varphi, \theta, \chi)$, shown in Fig. 1 using the Vector Model. $D_{KM}^J(\varphi, \theta, \chi)$ gives the probability amplitude of finding the principle axis of the molecule (the z' axis, parallel to the C-Cl bond and also the electric dipole moment of the molecule) at angle $\theta$ to the Z axis. The expectation value of $\cos\theta$ is given by $\langle \cos\theta \rangle = KM/J(J+1)$, so that selection of positive or negative values of $KM$ will preferentially orient the molecular principle axis parallel or antiparallel to the laboratory Z axis. Such $|JKM\rangle$ state selection has been used to study the dependence of photodissociation and bimolecular reactions on the orientation and alignment of symmetric-top molecules [2,3,4,5,6,7,8].

Second, the $Y_m^l(\theta, \varphi)$ spatial wavefunctions of orbital angular momentum are proportional to $D_{0\,m}^j(\varphi, \theta, \chi)$, with 0 projection of **l** along the z' axis (as $m' = 0$), so that **l** is always perpendicular to z. Therefore, the $D_{0\,m}^j(\varphi, \theta, \chi) \equiv Y_m^l(\theta, \varphi)$ describes the special case where z' is parallel to the position of the orbiting particle, and **l** is perpendicular to the orbital plane, as shown in Fig. 2a, using the Vector Model. This Vector-Model picture correctly shows that, asymptotically, the $Y_m^l(\theta, \varphi)$ and the probability of finding the particle will only have nonzero values within the classically-allowed angle ranges of $\frac{\pi}{2} - \theta_m \leq \theta \leq \frac{\pi}{2} + \theta_m$, where $\theta_m$ is the Vector-Model angle between **l** and the Z axis, given by $\cos\theta_m = m/|l|$. For finite values of $l$, the $Y_m^l(\theta, \varphi)$ will oscillate within this classically-allowed region, and will decay exponentially outside this region, in the classically-forbidden region, as shown in Fig. 2b.

Finally, for the third example, we will consider something unusual: choosing the projection $n = \sqrt{j(j+1)}$ in $D_{n\,m}^j(\varphi, \theta, \chi)$ so that the internal quantization axis z is parallel to the angular momentum $j$, as shown in Fig. 3 using the Vector Model (note that we now use z for the vector parallel to angular momentum $\boldsymbol{j}$, and reserve $z'$ for internal frames defined by other internal structure, such as the molecular dipole moment in Fig. 1). The asymptotic limit of this case was considered recently, where the asymptotic spatial wavefunction $D^{j}_{(j+\frac{1}{2})m}(\varphi,\theta,\chi)$ was used [9]:

$$D^{j}_{(j+\frac{1}{2})m}(\varphi,\theta,\chi) = e^{im\varphi}\,\delta(\theta - \theta_m)\,e^{i(j+\frac{1}{2})\chi} \tag{1}$$

where $\theta_m$ is the Vector-Model polar angle, given by $cos\theta_m = m/|\boldsymbol{j}|$. Note that in the high-$j$ limit, the asymptotic magnitude of $\boldsymbol{j}$ tends to $j + \frac{1}{2}$ (as $|\boldsymbol{j}| = \sqrt{j(j+1)} \to j + \frac{1}{2}$) so that $D^{j}_{(j+\frac{1}{2})m}(\varphi,\theta,\chi)$ is the asymptotic limit of $D^{j}_{\sqrt{j(j+1)},m}(\varphi,\theta,\chi)$. The spatial wavefunction of Eq. (1) allows the geometrical description of Clebsch-Gordan coefficients, Wigner rotation matrix elements $d^{j}_{m'm}(\varphi,\theta,\chi)$, m-state correlation matrix elements, and the calculation of the gyromagnetic ratio of elementary charged particles at the tree level ($g = 2$). These results are exact in the high-$j$ limit, but, surprisingly, they are either exact or an excellent approximation down to $j = \frac{1}{2}$ [9]. Clearly, it seems that an asymptotic spatial wavefunction that treats $\boldsymbol{j}$ as a three-dimensional entity is a useful concept. Based on the success of the unnormalizable wavefunction of Eq. (1), we ask whether the concept of an *exact* spatial wavefunction for $\boldsymbol{j}$ extends to finite values of $j$.

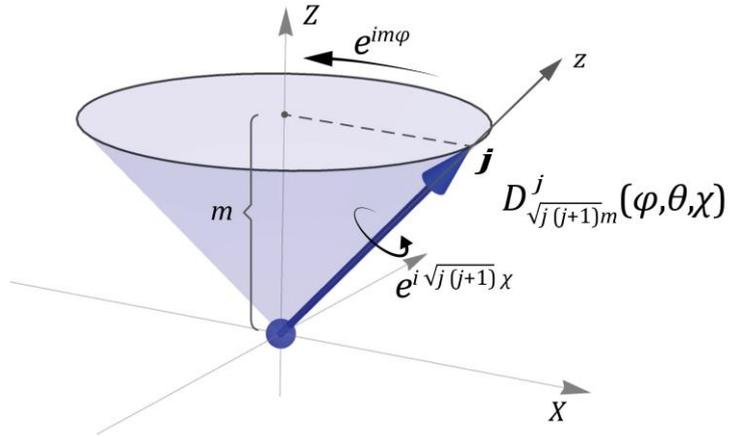

*Figure 3:* Depiction of the angular-momentum frame, where z is defined as being parallel to $\boldsymbol{j}$. In this frame, the spin-wavefunction is given by $D^{j}_{\sqrt{j(j+1)},m}(\varphi,\theta,\chi)$, for which the angular momentum j projects m along the space-fixed Z axis and $|\boldsymbol{j}|=\sqrt{j(j+1)}$ along z.

We note, however, that Eq. (1) has at least three unusual features:

(1) The internal projection quantum number $n$ seemingly violates the uncertainty principle, by taking a value $n = |\boldsymbol{j}| = j + \frac{1}{2}$ which is outside of the usual range $-j$ to $+j$.

(2) The quantum number $n$ assumes only a single value (a singlet: $n = j + \frac{1}{2}$) which is, therefore, non-rotatable to any other states in the internal coordinates, which is unusual for a non-zero angular momentum $j$.

(3) The wavefunction of Eq. (1) is only an asymptotic wavefunction for a reduced space defined by $\theta = \theta_m$, for appropriately reduced angular momentum operators [9].

Furthermore, the unusual features of Eq. (1) are not explained at a more fundamental level of quantum mechanical angular momentum, so that the origins of Eq. (1) are not made clear in [9]. One aim of this paper is to give a more fundamental explanation for Eq. (1), so that we can have a physical basis to extend Eq. (1) to low-$j$ spatial wavefunctions. To do so, we propose that angular momenta have a reduced, two-dimensional internal space, defined only by the operators $\hat{j}^2$ and $\hat{j}_z$. This assumption results in an internal projection $n = |\boldsymbol{j}| = \sqrt{j(j+1)}$ that is unique and constant, and $\boldsymbol{j}$ is non-rotatable in the internal coordinates, with the quantization axis $z$ always parallel to the angular momentum $\boldsymbol{j}$, as shown in Fig. 3 (explaining at a deeper level the three unusual features in the previous paragraph). We find that this assumption of a two-dimensional angular momentum in the internal coordinate allows us to reproduce the asymptotic spatial wavefunction of spin of Eq. (1) and to extend it to low $j$. At the end of the introduction, we describe the predictive power of these low-$j$ spatial wavefunctions of spin, which helps justify our proposal for this spatial wavefunction.

In the external $XYZ$ space $\hat{\boldsymbol{J}}^2$ is expressed in terms of $\hat{J}_X, \hat{J}_Y$, and $\hat{J}_Z$ as usual:

$$\hat{\boldsymbol{J}}^2 = \hat{J}_X^2 + \hat{J}_Y^2 + \hat{J}_Z^2 \tag{2a}$$

In contrast, for the internal reduced space of the two-dimensional angular momentum we have the simpler relation for the angular momentum operators $\hat{\boldsymbol{j}}$ and $\hat{j}_z$ [10]:

$$\hat{\boldsymbol{j}}^2 = \hat{j}_z^2 \tag{2b}$$

Finally, requiring that $\hat{\boldsymbol{J}}^2$ and $\hat{\boldsymbol{j}}^2$ have the same expectation value:

$$\langle \hat{\boldsymbol{J}}^2 \rangle = \langle \hat{\boldsymbol{j}}^2 \rangle = j(j+1) = \langle \hat{j}_z^2 \rangle \tag{3}$$

yields the expectation value of $\hat{j}_z$ to be $n = |\boldsymbol{j}| = \sqrt{j(j+1)} = \acute{j}$.

A consequence of the fact that $|n| > j$ is that the $D_{\acute{j}\,m}^{j}(\varphi, \theta, \chi)$ are not conventionally normalizable. The main aim of this paper is to show that the $D_{\acute{j}\,m}^{j}(\varphi, \theta, \chi)$ are useful wavefunctions, through a regularization procedure that allows the calculation of expectation values, which is an extension of a regularization procedure first introduced by Pandres for half-integer spherical harmonics [11]. To demonstrate the usefulness of the spatial-wavefunction formalism, we use the $D_{\acute{s},m}^{s}(\varphi, \theta, \chi)$ to compute directly the gyromagnetic ratio of elementary charged particles with spin, by calculating the expectation value of the magnetic-moment operator. Interestingly, g = 2 is computed for all values of spin, for $s > 0$. This agrees with the Dirac equation and Standard Model predictions for spins of ½ and 1, and confirms various arguments that $g = 2$ is the natural tree-level gyromagnetic ratio for all values of spin [12,13]. We note that, just as the magnitude of $\boldsymbol{j}$ is relativistically invariant [14], so is its maximum projection along itself $n = |\boldsymbol{j}| = \sqrt{j(j+1)}$.

Spatial wavefunctions are usually considered as problematic in relativistic quantum field theories, because of their seemingly non-covariant behavior under Lorentz transformations, and because they can spread faster than light. However, Pavšič has shown that a quantum field theoretic analysis, which distinguishes between basis position states and wave packet states, clarifies the issue of Lorentz covariance [15,16,17]. The issue of causality is resolved by observing that superluminal transmission of information cannot be achieved by such wave packets [16]. Therefore, spatial wavefunctions (such as those discussed here) are not clearly inconsistent with relativistic quantum field theory.

We give a brief outline of the paper. In Section II, we describe the wavefunctions $D^j_{n\,m}(\varphi,\theta,\chi)$ as solutions of the space-fixed angular momentum operators $\hat{J}^2, \hat{J}_Z, \hat{J}_\pm$, and of the body-fixed operator $\hat{j}_z$. In Section III, we introduce the definition of the inner product that allows the regularization of the $D^j_{n,m}(\varphi,\theta,\chi)$, and the calculation of the triple product of the $D^j_{n\,m}(\varphi,\theta,\chi)$ functions, needed for the calculation of expectation values (we use the term regularization, and avoid the term "renormalization", as this has a related but clearly different meaning in quantum electrodynamics). In Section IV we discuss the spatial distributions of the $D^j_{\jmath\,m}(\varphi,\theta,\chi)$ states, and how these connect with classical mechanics in the high-$j$ limit. In Section V we describe how the $D^j_{n,m}(\varphi,\theta,\chi)$ are consistent with the usual angular momentum coupling and rotational transformations of the $|jm\rangle$ states. In Section VI we calculate the $g$ factor of spins of any $S$ and $n$, from the expectation value of the magnetic moment operator. In Section VII we give arguments for why spin-1/2 particles have 3 internal orthogonal projections in three dimensional particles, along the $x', y', z'$ axes of the particle. In Section VIII we discuss how the spin wavefunctions of Standard Model particles $D^S_{n,m}(\varphi,\theta,\chi)$ fall into two categories: those with $n = s$ and those with $n = 0$ (for which the spin $s$ is perpendicular to the symmetry axis of the particle). We show that the internal-frame Clebsch-Gordan coefficients for allowed particle reactions are all unity, and thus do not affect the calculations of these reactions. We also present selection rules and a new conservation law, related to $n$, for particle reactions, from which we determine that the neutrino is a Dirac Fermion. In Section IX we discuss how the new degree of freedom of the internal projection $n$ opens the way for new dark-matter candidates (particularly those with $n = 0$), and we discuss one such possibility. In Section X we discuss some implications of the internal spin projections for isospin and weak isospin in the strong and the weak forces. Finally, in Section XI we discuss the conclusions of our results.

## II. WAVEFUNCTIONS

We propose the spatial wavefunction of the angular momentum $j$ to be the $D^j_{\jmath\,m}(\varphi,\theta,\chi)$ Wigner D-functions [where $\jmath = \sqrt{j(j+1)}$], which connect asymptotically to those of Eq. (1), to be:

$$D^j_{\jmath\,m}(\varphi,\theta,\chi) = e^{im\varphi}\, d^j_{\jmath\,m}(\theta)\, e^{i\jmath\chi} \qquad (4)$$

However, we will investigate the more general solution, $D^j_{n\,m}(\varphi,\theta,\chi)$, to the operator $\hat{J}^2$ in the external coordinates (Eq. 2a), for which the projection $n$ along $z$ can take any value:

$$D^j_{n\,m}(\varphi,\theta,\chi) = e^{im\varphi}\, d^j_{n\,m}(\theta)\, e^{in\chi} \tag{5}$$

where $m$ is the projection of the angular momentum $\boldsymbol{j}$ along the Z axis. The requirement that $D^j_{n\,m}(\varphi,\theta,\chi)$ be normalizable constrains $m$ and $n$ to range from $-j$ to $+j$, in integer steps, as is the case for the well-known Wigner D-functions $D^j_{m'\,m}(\varphi,\theta,\chi)$. For the $D^j_{n\,m}(\varphi,\theta,\chi)$ we use the more general body-fixed projection $n$, because in this Section we will treat $n$ as a continuous variable, for values outside of the normal range of $-j$ to $+j$, to better understand the properties of the $D^j_{n\,m}(\varphi,\theta,\chi)$.

The $D^j_{n\,m}(\varphi,\theta,\chi)$ wavefunctions are clearly eigenfunctions of the space-fixed angular momentum operator $\hat{J}_Z$, given by (henceforth we set $\hbar = 1$) [18]:

$$\hat{J}_Z = -i\frac{\partial}{\partial \varphi} \tag{6}$$

yielding eigenvalue $m$, corresponding to the projection of $\boldsymbol{j}$ along the Z.

The $D^j_{n\,m}(\varphi,\theta,\chi)$ wavefunctions are also eigenfunctions of the body-fixed angular momentum operator $\hat{j}_z$, given by [18]:

$$\hat{j}_z = -i\frac{\partial}{\partial \chi} \tag{7}$$

yielding the eigenvalue $n$.

We will now determine the angular dependence of the $D^j_{n\,m}(\varphi,\theta,\chi)$, which must be eigenfunctions of the $\hat{J}^2$ operator, given by [18]:

$$\hat{J}^2 = -\left\{\frac{\partial^2}{\partial \theta^2} + \cot\theta\,\frac{\partial}{\partial \theta} + \frac{1}{\sin^2\theta}\left(\frac{\partial^2}{\partial \varphi^2} + \frac{\partial^2}{\partial \chi^2} - 2\cos\theta\,\frac{\partial^2}{\partial \varphi \partial \chi}\right)\right\} \tag{8}$$

Operating $\hat{J}^2$ on $D^j_{n\,m}(\varphi,\theta,\chi)$ in Eq. (5a) yields the eigenvalue $j(j+1)$, and evaluating the derivatives in $\varphi$ and $\chi$ gives a differential equation in $\theta$ for $d^j_{n\,m}(\theta)$:

$$\left[\frac{\partial^2}{\partial \theta^2} + \cot\theta\,\frac{\partial}{\partial \theta} - \frac{1}{\sin^2\theta}(m^2 + n^2 - 2mn\cos\theta) + j(j+1)\right] d^j_{n\,m}(\theta) = 0 \tag{9}$$

The general solution for $d^j_{n\,m}(\theta)$ is given by:

$$d^j_{n\,m}(\theta) = \left[A\left(\sin\frac{\theta}{2}\right)^{m-n}\left(\cos\frac{\theta}{2}\right)^{m+n} {}_2F_1\left(m-j, m+j+1; 1+m-n; \sin^2\frac{\theta}{2}\right) + \right.$$
$$\left. B\left(\sin\frac{\theta}{2}\right)^{n-m}\left(\cos\frac{\theta}{2}\right)^{n+m} {}_2F_1\left(n-j, n+j+1; 1+n-m; \sin^2\frac{\theta}{2}\right)\right] \tag{10}$$

where ${}_2F_1(a,b;c;z)$ is the Gaussian hypergeometric function, and $A$ and $B$ are constants. Note that for $n = m'$ (where $m'$ ranges from $-j$ to $j$ in integer steps), Eq. (10) can be reduced to the conventional Wigner $d^j_{m'\,m}(\theta)$ functions.

The operator $\hat{J}_\pm$ raises or lowers the $m$ quantum number:

$$\hat{J}_\pm D^j_{n\,m}(\varphi,\theta,\chi) = \sqrt{j(j+1) - m(m\pm 1)}\, D^j_{n,m\pm 1}(\varphi,\theta,\chi) \tag{11}$$

and is given by [18]:

$$\hat{J}_\pm = e^{\pm i\varphi}\left\{i\left[\cot\theta\frac{\partial}{\partial\varphi} - \frac{1}{\sin\theta}\frac{\partial}{\partial\chi}\right] \pm \frac{\partial}{\partial\theta}\right\} \tag{12}$$

An apparent exception to the standard rule of Eq. (11) are the maximum projection states, where a non-zero function seems to be given, with projection $m = \pm(j+1)$, beyond the physical range:

$$\hat{J}_\pm D^j_{n,\pm j}(\varphi,\theta,\chi) = D^j_{n,\pm(j+1)}(\varphi,\theta,\chi) \tag{13}$$

However, using work by Pandres and others [11,19,20], it can be demonstrated that the states $D^j_{n,\pm(j+1)}(\varphi,\theta,\chi)$ are vanishing functions, as shown by two facts: (1) lowering/raising them back to $m = \pm j$ yields zero (see Appendix A):

$$\hat{J}_\mp D^j_{n,\pm(j+1)}(\varphi,\theta,\chi) = 0 \tag{14}$$

and (2) their norm, using the inner project defined in Section III, also vanish:

$$\left\langle D^j_{n,\pm(j+1)}(\varphi,\theta,\chi)\,\middle|\, D^j_{n,\pm(j+1)}(\varphi,\theta,\chi)\right\rangle = 0 \tag{15}$$

Using the general solution of Eq. (10b), we give expressions for the $d^j_{n\,m}(\theta)$ for spin 0, ½, and 1; specifically, we give expressions for spin 0 (which is merely a constant):

$$d^0_{n,0}(\theta) = N_0 \tag{16}$$

for spin ½:

$$d^{\frac{1}{2}}_{n,1/2}(\theta) = N_{1/2}\left[e^{\frac{i\pi}{4}}\left(\cos\frac{\theta}{2}\right)^{\frac{1}{2}+n}\left(\sin\frac{\theta}{2}\right)^{\frac{1}{2}-n} + e^{-\frac{i\pi}{4}}\left(\cos\frac{\theta}{2}\right)^{-\frac{1}{2}-n}\left(\sin\frac{\theta}{2}\right)^{n-\frac{1}{2}}\left(n+\frac{1}{2}\cos\theta\right)\right] \tag{17a}$$

$$d^{\frac{1}{2}}_{n,-1/2}(\theta) = N_{1/2}\left[e^{\frac{i\pi}{4}}\left(\cos\frac{\theta}{2}\right)^{n-\frac{1}{2}}\left(\sin\frac{\theta}{2}\right)^{-\frac{1}{2}-n}\left(n-\frac{1}{2}\cos\theta\right) + e^{-\frac{i\pi}{4}}\left(\cos\frac{\theta}{2}\right)^{\frac{1}{2}-n}\left(\sin\frac{\theta}{2}\right)^{\frac{1}{2}+n}\right] \tag{17b}$$

and for spin 1:

$$d^1_{n,1}(\theta) = N_1\left[e^{\frac{i\pi}{4}}\left(\cos\frac{\theta}{2}\right)^{1+n}\left(\sin\frac{\theta}{2}\right)^{1-n} + \frac{e^{-\frac{i\pi}{4}}\left(\cos\frac{\theta}{2}\right)^{-n-1}\left(\sin\frac{\theta}{2}\right)^{n-1}\left(n^2+n\cos\theta-\frac{1}{2}\sin^2\theta\right)}{2}\right] \tag{18a}$$

$$d^1_{n,0}(\theta) = N_1\left[\frac{e^{\frac{i\pi}{4}}}{\sqrt{2}}\left(\cos\frac{\theta}{2}\right)^n\left(\sin\frac{\theta}{2}\right)^{-n}(n-\cos\theta) + \frac{e^{-\frac{i\pi}{4}}}{\sqrt{2}}\left(\cos\frac{\theta}{2}\right)^{-n}\left(\sin\frac{\theta}{2}\right)^n(n+\cos\theta)\right] \tag{18b}$$

$$d^1_{n,-1}(\theta) = N_1\left[\frac{e^{\frac{i\pi}{4}}\left(\cos\frac{\theta}{2}\right)^{n-1}\left(\sin\frac{\theta}{2}\right)^{-n-1}\left(n^2-n\cos\theta-\frac{1}{2}\sin^2\theta\right)}{2} + e^{-\frac{i\pi}{4}}\left(\cos\frac{\theta}{2}\right)^{1-n}\left(\sin\frac{\theta}{2}\right)^{1+n}\right] \tag{18c}$$

where the $N_j$ are normalization constants, and the constants $A$ and $B$ in Eq. (10b) have been chosen so that the $(2j+1)$-degenerate group of $D^j_{n,m}(\varphi,\theta,\chi)$ satisfy Eqs. (11-14) and are geometrically symmetric: lowering/raising from $\pm m$ to $\mp m$ corresponds to the geometrical transformation $\theta \to \pi - \theta$ and complex conjugation, which gives $\cos\frac{\theta}{2} \to \sin\frac{\theta}{2}$, $\sin\frac{\theta}{2} \to \cos\frac{\theta}{2}$, $\cos\theta \to -\cos\theta$, and sign reversal of the phase factors $e^{\pm im\varphi} \to e^{\mp im\varphi}$ and $e^{\pm\frac{i\pi}{4}} \to e^{\mp\frac{i\pi}{4}}$. Notice that for this geometrical transformation, $[d^j_{n,\pm m}(\varphi, \pi - \theta, \chi)]^* = d^j_{n,\mp m}(\varphi, \theta, \chi)$. In addition, notice that for the geometrical transformations $\theta \to \theta + 2\pi$ or $\varphi \to \varphi + 2\pi$, that $D^j_{n,m}(\varphi, \theta, \chi) \to (-1)^{2j} D^j_{n,m}(\varphi, \theta, \chi)$, so that functions with half-integer $j$ change sign, whereas those with integer $j$ do not.

Wavefunctions for any $j$ can be generated by setting the first term of the $d^j_{n,j}$ state as $e^{\frac{i\pi}{4}}\left(\cos\frac{\theta}{2}\right)^{j+n}\left(\sin\frac{\theta}{2}\right)^{j-n}$; subsequently all the first terms of the states down to $m = -j$ are produced by operating $\hat{J}_-$ sequentially on $D^j_{n,j}$. Similarly, the second term of the $d^j_{n,-j}$ state is set as $e^{-\frac{i\pi}{4}}\left(\sin\frac{\theta}{2}\right)^{j+n}\left(\cos\frac{\theta}{2}\right)^{j-n}$, and the second terms of the states up to $m = +j$ are produced by operating $\hat{J}_+$ sequentially on $D^j_{n,-j}$. In both cases, after each operation of $\hat{J}_\pm$, the result must be divided by the factor $\sqrt{j(j+1) - m(m\pm 1)}$, from Eq. (11).

### III. INNER PRODUCT REGULARIZATION AND EXPECTATION VALUES

The expressions of Eqs. (17-18) are unnormalizable for $n = \sqrt{j(j+1)} = \dot{j}$, because at least one term in each expression gives a divergent integral. In fact, in general, all terms for all values of $j$ are divergent, except for one term of the maximum projection states. To regularize these expressions, we use an extension of the definition of the inner product $\langle\Psi|\Psi'\rangle$ given by Pandres [11] for half-integer spherical harmonics (to include an integral over the angle $\chi$, over an infinite domain):

$$\langle\Psi|\Psi'\rangle = \int_0^\pi d\theta \left[\sin\theta \int_0^{2\pi} d\varphi \int_0^{2\pi q} \lim_{q\to\infty} \frac{1}{2\pi q} d\chi\, \Psi^*(\varphi,\theta,\chi)\Psi'(\varphi,\theta,\chi) - f(\theta)\right] \quad (19a)$$

where $f(\theta)$ is:

$$f(\theta) = \sum_{m=1}^\infty a_m(\sin\theta)^{-m} + \cos\theta \sum_{m=1}^\infty b_m(\sin\theta)^{-m}, \quad (19b)$$

and the constants $a_m$ and $b_m$ are chosen so that diverging terms in $\Psi^*\Psi' \sin\theta$, which are propotional to negative powers of $\sin\theta$, are cancelled. Note that $\Psi^*\Psi' \sin\theta$ is first expanded in powers of $\sin\theta$, to separate terms with negative powers (that diverge) from non-negative powers (that don't diverge). We interpret the extension of the regularization of Pandres [11] of Eq. (19) to be the correct inner product for spatial wave functions of angular momenta which are three-dimensional in the external frame and two-dimensional in the internal frame. The divergent terms subtracted using Eq. (19b) are the terms needed to remove the consequences of the angular uncertainty principle in the internal frame from $[\hat{j}_x, \hat{j}_y] = i\hat{j}_z$, which forbids normalizable wavefunctions with projection $m > j$. However, the angular uncertainty principle does not hold in the internal frame as there are no $\hat{j}_x$ or $\hat{j}_y$ operators. Therefore, Equation (19) allows the regularization of the $D^j_{n,m}(\varphi, \theta, \chi)$ for $n = \jmath$, with $\boldsymbol{j}$ parallel to $z$. In fact, it regularizes the $D^j_{n,m}(\varphi, \theta, \chi)$ for all $n$, from $-\infty < n < \infty$. It may seem strange that Eq. (19) regularizes $D^j_{n,m}(\varphi, \theta, \chi)$ for all $n$, even though Standard-Model particles only require values $n = \jmath$ or $n = 0$ (see Section VIII). However, this larger range of $n$ is needed, when considering angular momentum coupling, to describe the projections of angular momenta onto the internal symmetry axes of the other angular momenta (as described in Section VIII and Appendix C).

The actual value of the normalization constant $N_j$ is not relevant for two reasons: (1) the unnormalizable state cannot be used to calculate the direct angular distribution of $\boldsymbol{j}$, as only expectation values can be calculated (see Section IV), and (2) expectation values are expressed as a ratio where $N_j$ cancels.

The inner product can be used to show analytically that:

$$\left\langle D^j_{n',m'}(\varphi, \theta, \chi) \mid D^j_{n,m}(\varphi, \theta, \chi) \right\rangle = \delta_{n'n}\delta_{m'm} \tag{20}$$

for $-j \leq n \leq +j$ without regularization, as these wavefunctions are normalizable. For $|n| > j$ Eq. (19) will regularize the inner product for all $n$.

The integral of the triple product of $D^j_{m' m}(R)$ functions is given by [18]:

$$\left\langle D^{j_1}_{m'_1 m_1}(R) D^{j_2}_{m'_2 m_2}(R) \mid D^{j_3}_{m'_3 m_3}(R) \right\rangle = 8\pi^2 \begin{pmatrix} j_1 & j_2 & j_3 \\ -m'_1 & -m'_2 & m'_3 \end{pmatrix} \begin{pmatrix} j_1 & j_2 & j_3 \\ -m_1 & -m_2 & m_3 \end{pmatrix} \tag{21a}$$

Importantly, Eq. (19) allows the extension of the integral of Eq. (21a), of the triple product of $D^j_{m' m}(R)$ functions, to those of $D^j_{n m}(R)$ functions:

$$\left\langle D^{j_1}_{n_1 m_1}(R) D^{j_2}_{n_2 m_2}(R) \mid D^{j_3}_{n_3 m_3}(R) \right\rangle = 8\pi^2 \begin{pmatrix} j_1 & j_2 & j_3 \\ -n'_1 & n'_1 - n_3 & n_3 \end{pmatrix} \begin{pmatrix} j_1 & j_2 & j_3 \\ -m_1 & -m_2 & m_3 \end{pmatrix} \tag{21b}$$

where $n'_1$ and $n'_2 = n'_1 - n_3$ are the projections of $\boldsymbol{j}_1$ and $\boldsymbol{j}_2$ along the $z$ axis of $\boldsymbol{j}_3$, so that $n'_1 + n'_2 = n_3$, and $m_1 + m_2 = m_3$. Equation (21b) is derived as is Eq. (21a) [18], with the difference that sums over $m'$ are replaced by integrals over $n$, and that the orthogonality relation from Eq. (19) is used instead of the normal orthogonality relation for $D^j_{m',m}(\varphi, \theta, \chi)$. Equation (21b) can be verified analytically for any value of $n_i$ in the range $-j_i \leq n_i \leq +j_i$ for both terms in the $D^{j_i}_{n_i m_i}(R)$, and for $|n_i| > j_i$ for the first term of each $D^{j_i}_{n_i j_i}(R)$, which is not divergent. For the divergent terms of the $D^{j_i}_{n_i m_i}(R)$, Eq. (21b) can be verified straightforwardly for interger $n_i > j_i$. Equation (21b) allows the calculation of expectation values of $D^j_{n m}(R)$ wavefunctions for any $n$. In Appendix B, we demonstrate a calculation of the expectation value of (a) $cos\theta$ for $j = m = 1/2$, and $n = 1$, and (b) $P_2(cos\theta)$ for $j = m = 1$, and $n = 3/2$, demonstrating the regularization of Eq. 19 and the use of Eq. (21b). For $D^{j_2}_{n_2 m_2}(R) = D^k_{0 q}(R)$, the $D^k_{0\,q}(R)$ are proportional to the spherical harmonics $Y^k_q(\theta,\varphi)$, which span the space of all angular functions, so that the effect of the angular part of any operator on $D^j_{n\,m}(R)$ can be calculated with Eq. (21b).

We give a special case of Eq. (21b) below, which, along with Eq. (21b), is a key result of this paper. For two of the D-functions with the same value of $j$, $n$, and $m$, and the third, $D^k_{0\,0}(R)$, with $n = m = 0$, we have the expectation value $\langle D^k_{0\,0}(R)\rangle$:

$$\frac{\langle D^j_{n\,m}(R) D^k_{0\,0}(R) | D^j_{n\,m}(R)\rangle}{\langle D^j_{n\,m}(R) D^0_{0\,0}(R) | D^j_{n\,m}(R)\rangle} = \frac{\begin{pmatrix} j & k & j \\ n & 0 & n \end{pmatrix}\begin{pmatrix} j & k & j \\ m & 0 & m \end{pmatrix}}{\begin{pmatrix} j & 0 & j \\ n & 0 & n \end{pmatrix}\begin{pmatrix} j & 0 & j \\ m & 0 & m \end{pmatrix}} = \langle j\,n, k\,0 | j\,n\rangle \langle j\,m, k\,0 | j\,m\rangle \quad (22)$$

Equation (22) is used for the calculation of expectation values, such as for the calculation of the gyromagnetic ratio $g = 2$, from the calculation of the expectation value of the magnetic moment operator. Equation (21b) or (22) can also be used to calculate selection rules in particle reactions, from the overlap integral of the spin wavefunctions (see Section VIII).

Clebsch-Gordan coefficients in the internal frame are not, in all cases, calculated as those in the external frame. The main reason for this is that, in the external frame, all angular momenta have a common quantization axis $Z$, whereas in the internal frame, each angular momentum has a separate internal axis $z$. However, the internal Clebsch-Gordan coefficients of the form $\langle j\,n, k\,0 | j\,n\rangle$ share a common internal $z$ axis, and therefore these can be calculated using the standard formulas. In particular, the $\langle j\,n, k\,0 | j\,n\rangle$ have analytical expressions, which are mathematically valid for any value of $n$; for example, we give the values for $k$ ranging from 0 to 4:

$$\langle j\,n, 0\,0 | j\,n\rangle = P_0(\cos\theta_n) = 1 \quad (23a)$$

$$\langle j\,n, 1\,0 | j\,n\rangle = P_1(\cos\theta_n) = \frac{n}{\sqrt{j(j+1)}} \quad (23b)$$

$$\langle j\,n, 2\,0 | j\,n\rangle = \sqrt{\frac{j(j+1)}{(j-1/2)(j+3/2)}}\, P_2(\cos\theta_n) \quad (23c)$$

$$\langle j\,n, 3\,0 | j\,n\rangle = \frac{j(j+1)}{\sqrt{(j-1)(j-1/2)(j+3/2)(j+2)}} \left[ P_3(\cos\theta_n) + \frac{\cos\theta_n}{2j(j+1)} \right] \quad (23d)$$

$$\langle j\,n, 4\,0 | j\,n\rangle = \sqrt{\frac{j^3(j+1)^3}{(j-3/2)(j-1)(j-1/2)(j+3/2)(j+2)(j+5/2)}} \left[ P_4(\cos\theta_n) + \frac{25\cos^2\theta_n - 6}{8j(j+1)} \right] \quad (23e)$$

where $\cos\theta_n = n/\sqrt{j(j+1)}$, and the $P_k(x)$ are Legendre polynomials of rank $k$.

In Section VIII (and Appendix C) we demonstrate how Clebsch-Gordan coefficients $\langle j_1\,n'_1, j_2\,n'_2 | j_3\,n_3\rangle$ in the internal frame are calculated, that don't have the form $\langle j\,n, k\,0 | j\,n\rangle$, and how these are relevant to Standard-Model particle reactions. In the next section, we use Eq. (19) to examine the angular distribution of orbital angular momentum.

## IV. ANGULAR MOMENTUM SPATIAL DISTRIBUTIONS

The spatial distribution of a spatial wavefunction with $n = j$ is given by the square $\left|D^j_{j\,m}(\varphi,\theta,\chi)\right|^2$. However, with the exception of the trivial case of the $j = 0$ state (where $\left|D^0_{0,0}(\varphi,\theta,\chi)\right|^2 = N_0^2$ is constant with no dependence on angles), these squares are unnormalizable, so that the angular distribution of $j$ does not correspond to an observable.

However, the magnetic moment $\mu$ of $j$ can be probed with a magnetic field $\mathbf{B}$, through the interaction Hamiltonian $H = -\mu \cdot \mathbf{B} = |\mu||\mathbf{B}|\cos\theta$. The expectation value of $\cos\theta = D^1_{0,0}(0,\theta,0)$ can be calculated using Eq. (22):

$$\langle \cos\theta \rangle = \langle j\,j, 1\,0|j\,j\rangle \langle j\,m, k\,0|j\,m\rangle \tag{24a}$$

where $\langle j\,j, 1\,0|j\,j\rangle = 1$ from Eq. (22a), and thus:

$$\langle \cos\theta \rangle = \frac{m}{\sqrt{j(j+1)}} = \cos\theta_m \tag{24b}$$

which agrees with known quantum mechanics and the Vector Model.

We then investigate the limit $j \to \infty$, and the calculation of $\langle P_k(\cos\theta)\rangle$ for all $k$.

The Clebsch-Gordan coefficient $\langle j\,j, k\,0|j\,j\rangle$ can be written in the form:

$$\langle j\,j, k\,0|j\,j\rangle = U_k(j)\left[P_k(\cos\theta) + \frac{f(\theta)}{j(j+1)}\right] \tag{25}$$

where $U_k(j)$ is a constant, and $f(\theta)$ is proportional to powers of $\cos\theta$ up to rank $(k-2)$. In the limit of $j \to \infty$, $U_k(j) \to 1$ and $f(\theta)/j(j+1) \to 0$, so that $\langle j\,j, k\,0|j\,j\rangle \to 1$, and the expectation values $\langle P_k(\cos\theta)\rangle$ become well defined. Specifically, in this limit, $\langle j\,m, k\,0|j\,m\rangle \to P_k(\cos\theta_m)$, so that:

$$\langle P_k(\cos\theta)\rangle \to P_k(\cos\theta_m) \quad \text{for } j \to \infty \tag{26}$$

where $\cos\theta_m = m/\sqrt{j(j+1)}$. This limit of the expectation value of Eq. (26) is consistent with the classical limit of the Vector Model:

$$\left|D^j_{j\,m}(\varphi,\theta,\chi)\right|^2 \to \delta(\theta - \theta_m) \quad \text{for } j \to \infty \tag{27}$$

We conclude that the expectation values $\langle P_k(\cos\theta)\rangle$ agree with known quantum mechanics: for $k = 1$ $\langle \cos\theta\rangle$ gives the well-known projection of $j$ along the Z axis, in Eq. (24). In the classical limit of $j \to \infty$, the $\langle P_k(\cos\theta)\rangle$ and $\left|D^j_{j\,m}(\varphi,\theta,\chi)\right|^2$ give values consistent with the classical limit of the Vector Model, given by Eqs. (26) and (27), and the full spatial distribution of $j$ becomes consistent with the semiclassical limit (see also the discussion of angular momentum wavepackets in [9]).

## V. CLEBSCH-GORDAN COEFFICIENTS AND WIGNER D-FUNCTIONS IN THE EXTERNAL FRAME

*1. Clebsch-Gordan coefficients.* For the inner product in Eq. (19), $\hat{J}_+$ and $\hat{J}_-$ are mutual Hermitian adjoints (proved similarly as in the Appendix of [11]):

$$\langle \hat{J}_+ \Psi | \Psi' \rangle = \langle \Psi | \hat{J}_- \Psi' \rangle \tag{28}$$

Using Eq. (28), it is straightforward to show that [11]:

$$\left\langle D^j_{n\,m}(\varphi,\theta,\chi) \,\middle|\, D^{j'}_{n'\,m'}(\varphi,\theta,\chi) \right\rangle = \delta_{jj'}\delta_{nn'}\delta_{mm'} \qquad (for\ |m|, |m'| \leq j) \tag{29a}$$

$$\left\langle D^j_{n\,m}(\varphi,\theta,\chi) \,\middle|\, D^{j'}_{n'\,m'}(\varphi,\theta,\chi) \right\rangle = 0 \qquad (for\ |m|\ or\ |m'| > j) \tag{29b}$$

As the $D^j_{n\,m}$ functions form an orthonormal set [examples given in Eqs. (17-18)] and obey Eqs. (8) and (11), all Clebsch-Gordan coefficients can be calculated in the usual way using the lowering operator $\hat{J}_-$ [18,21].

*2. Wigner D-functions for space-fixed-axes rotations.* The usual Wigner rotation matrix elements $d^{j_1}_{m'_1 m_1}(\theta)$ [18,21,22], for rotations in the space-fixed axis, can be expressed in terms of a Clebsch-Gordan coefficient, where the $|j_1 m_1\rangle$ state is coupled to an infinite angular momentum $j_2$ at angle $\theta$ to the Z axis [9]:

$$d^{j_1}_{m'_1 m_1}(\theta) = (-1)^{j_1 - m'_1} \lim_{j_2 \to \infty} \langle j_1 m_1, j_2 m_2 \,|\, j_2 + m'_1, m_1 + m_2 \rangle \tag{30}$$

where $\cos\theta = m_2/j_2$; note that $m_2 = j_2 \cos\theta$ also tends to infinity. This coupled state decouples uniquely to the $|j_1\, m'_1\rangle$ state along the z quantization axis (parallel to $\boldsymbol{j}_2$):

$$\lim_{j_2 \to \infty} \langle j_2 + m'_1, j_2 + m'_1 \,|\, j_2\, j_2, j_1\, m'_1 \rangle = 1 \tag{31}$$

Therefore, Eqs. (30-31) describe the probability amplitude that the $|j_1 m_1\rangle$ state, along Z, is projected to the $|j_1\, m'_1\rangle$ state along a new quantization axis z, at angle $\theta$ to Z, which is the definition of $d^{j_1}_{m'_1 m_1}(\theta)$. Therefore, Eq. (30) shows that the $D^j_{n\,m}$ functions transform under rotation as the $|jm\rangle$ state.

We demonstrate this for $S = 1/2$, using the general Clebsch-Gordan coefficient for the $\left|\frac{1}{2}\frac{1}{2}\right\rangle$ state [18]:

$$\left\langle \tfrac{1}{2}\tfrac{1}{2}, j_2\, m_2 \,\middle|\, j_2 \pm \tfrac{1}{2}, m_2 + \tfrac{1}{2} \right\rangle = \sqrt{\frac{j_2 \pm m_2 + 1}{2 j_2 + 1}} \tag{32}$$

Inserting Eq. (32) into Eq. (30) yields, for $m'_1 = \tfrac{1}{2}$ and $-\tfrac{1}{2}$, and using $m_2 = j_2 \cos\theta$:

$$d^{\tfrac{1}{2}}_{\tfrac{1}{2}\tfrac{1}{2}}(\theta) = d^{\tfrac{1}{2}}_{-\tfrac{1}{2}-\tfrac{1}{2}}(\theta) = \lim_{j_2 \to \infty} \sqrt{\frac{j_2(1+\cos\theta)+1}{2 j_2 + 1}} = \sqrt{\frac{(1+\cos\theta)}{2}} = \cos\frac{\theta}{2} \tag{33a}$$

$$d^{\tfrac{1}{2}}_{-\tfrac{1}{2}\tfrac{1}{2}}(\theta) = -d^{\tfrac{1}{2}}_{\tfrac{1}{2}-\tfrac{1}{2}}(\theta) = \lim_{j_2 \to \infty} \sqrt{\frac{j_2(1-\cos\theta)+1}{2 j_2 + 1}} = \sqrt{\frac{(1-\cos\theta)}{2}} = \sin\frac{\theta}{2} \tag{33b}$$

showing, as expected, that the $D^{\frac{1}{2}}_{n\,m}(\varphi,\theta,\chi)$ transform under rotation as spinors.

*3. Wigner D-functions for body-fixed-axes rotations.* We defined the body-fixed $z$ axis to be exactly parallel to the angular momentum $\boldsymbol{j}$ (or parallel to the internal body-fixed symmetry axis of an elementary particle). This definition seems to preclude any body-fixed-axis rotation of the $z$ axis, as it is fixed with respect to $\boldsymbol{j}$ (or the body-fixed axis), by definition. We examine the internal consistency of the angular momentum algebra, by considering the Wigner rotation matrix $d^j_{n,n}(\theta)$, which gives the probability amplitude for the rotation of the $z$ axis by angle $\theta$. The $d^j_{n,n}(\theta)$ function is unnormalizable for $n = \sqrt{j(j+1)}$, however, we can calculate the expectation value of $\cos\theta$, using Eq. (22):

$$\langle \cos\theta \rangle = \langle j\,n, 1\,0 | j\,n \rangle^2 = \frac{n^2}{j(j+1)} \tag{34}$$

where $\langle j\,n, 1\,0 | j\,n \rangle$ is given by Eq. (23b). For $n = \sqrt{j(j+1)}$, Eq. (34) gives that $\langle \cos\theta \rangle = 1$, which can be interpreted by the fact that the regularized $\left|d^j_{n,n}(\theta)\right|^2$ distribution is singularly distributed at $\theta = 0$, equivalent to $\delta(\theta)$. We conclude that the $D^j_{n\,m}(\varphi,\theta,\chi)$ with $n = \sqrt{j(j+1)} = \hat{\jmath}$ cannot be rotated to change the projection $n$. This argument applies to all three of the orthogonal states described in Section VII, as they each correspond to a maximum projection state: $n_{z'} = \sqrt{j(j+1)}$, or $n_{x'} = \sqrt{j(j+1)}$, or $n_{y'} = \sqrt{j(j+1)}$. The symmetry of each non-rotatable state is U(1).

## VI. G-FACTOR CALCULATION

We now use the spin wavefunctions to calculate the $g$ factor of an elementary particle, which has mass $M$, charge $e$, spin $s$, and a spatial angular momentum wavefunction $\psi_{spin}(\varphi,\theta,\chi) = D^s_{n\,m}(\varphi,\theta,\chi)$, given by Eq. (5), with $s = j$. It will have magnetic moment components $\hat{\mu}_Z$ and $\hat{\mu}_z$, along the $Z$ and $z$ axes, respectively, given by:

$$\hat{\mu}_Z = \mu\,\hat{S}_Z = -i\mu\frac{\partial}{\partial\varphi} \tag{35a}$$

$$\hat{\mu}_z = \mu\,\hat{S}_z = -i\mu\frac{\partial}{\partial\chi} \tag{35b}$$

where $\mu = e\hbar/(2M)$. The $z$ axis is at angle $\theta$ to $Z$, and $z$ is distributed about $Z$ with cylindrical symmetry, so that only the parallel component of $\mu_z$ contributes to the total magnetic moment along $Z$:

$$\hat{\mu}^T_Z = \hat{\mu}_Z + \cos\theta\,\hat{\mu}_z \tag{36}$$

The magnetic moment is then given by the expectation value of $\hat{\mu}^T_Z$ for the state $D^s_{n\,m}(\varphi,\theta,\chi) \equiv D^s_{n\,m}(R)$, given by:

$$\langle \hat{\mu}^T_Z \rangle = \frac{\langle D^s_{n\,m}(R) | (\hat{\mu}_Z + \cos\theta\,\hat{\mu}_z) D^s_{n\,m}(R) \rangle}{\langle D^s_{n\,m}(R) | D^s_{n\,m}(R) \rangle} = \mu\,(m + n\langle \cos\theta\rangle) = \mu\,m\left(1 + \frac{n^2}{s(s+1)}\right) \tag{37}$$

where the expectation value $\langle \cos\theta \rangle = nm/[s(s+1)]$ is given by Eqs. (22) and (23b). The magnetic moment of an elementary particle is expressed generally as $g\mu m$, so that we find in general:

$$g = \left(1 + \frac{n^2}{s(s+1)}\right) \tag{38}$$

If we require that $g = 2$ (at the tree level) for charged particles independent of the value of $S$ [14], then Eq. (38) gives a clear geometrical interpretation (and without relativistic considerations [21,23]): we see that the projection of $\boldsymbol{s}$ along the body-fixed z axis must be maximal, $n = \sqrt{s(s+1)}$, and a magnetic moment points along $\boldsymbol{s}$ with magnitude $\mu_B\sqrt{s(s+1)}$. This result supports the picture of the two-dimensional angular momentum shown in Fig. 3. In addition, $\hat{\mu}_Z$ and $\hat{\mu}_z$ contribute equally to the magnetic moment of the particle along Z.

We note that this straightforward geometric calculation of $g = 2$ for all $s$ (given $n = \sqrt{s(s+1)}$) cannot be calculated straightforwardly in standard quantum mechanics for $s > 1$ [12,13]. The interaction of a charged particle with spin $s = 1/2$ with an electromagnetic field can be taken into account by substituting the derivative of charged fields in the Lagrangian with the covariant derivative [24]:

$$\partial_\mu \varphi \to \partial_\mu \varphi + ieA_\mu \tag{39}$$

where $e$ is the electric charge of the field $\varphi$, and the $A_\mu$ are the components of the magnetic vector potential. The substitution of Eq. (39) is known as the "minimal substitution" or "minimal coupling", and it correctly yields $g = 2$ for $s = 1/2$. Using "minimal coupling" in the Proca equation for $S = 1$ [25,26] yields $g = 1$. Belinfante then calculated $g = 2/3$ for $s = 3/2$, and subsequently proposed the Belinfante conjecture [27]: that $g = 1/S$, for all spins $S$. However, the charged W boson with $S = 1$ is observed to have $g = 2$ [28], experimentally refuting Belifante's conjecture, and showing that "minimal coupling" is not sufficient; the requirement that the Lagrangian be gauge invariant adds an extra term to the Lagrangian [29], which then gives $g = 2$.

Several arguments are given elsewhere [12,13] that $g = 2$ is the natural value for all elementary charged particles with any spin $S$. Here, we add an additional geometric argument: that $g = 2$ for any spin $S$ corresponds to $\boldsymbol{S}$ and the magnetic moment $\boldsymbol{\mu}$ being parallel or antiparallel to z, with $n = \sqrt{s(s+1)}$, as shown in Fig. 3, where z defines the body-fixed axis of the $D^s_{n\,m}(\varphi,\theta,\chi)$ symmetric-top wavefunction. In contrast, solving Eq. (38) for the Belifante conjecture of $g = 1/s$ yields $n = \sqrt{(1-s^2)}$. This result, although it gives the correct result for $s = 1/2$ (that $n = \sqrt{3}/2 = \sqrt{s(s+1)}$), gives the physically meaningless result that $n$ is imaginary for $s > 1$.

## VII. THE THREE INTERNAL SPIN PROJECTIONS FOR SPIN-1/2

From Section VI we see that all charged fundamental particles with spin $s$ (i.e., the quarks, electrons, and the W boson) must have internal projection $n = \sqrt{s(s+1)} = s$. However, we will see in the next section that neutral particles that are their own antiparticle must have the internal projection $n_z = 0$ (e.g., for the photon and the Z boson, and also for the gluon), where $z'$ is an internal symmetry axis of the particle. Spin wave functions with $n = 0$ have already been described by Pandres [11,19], in the description of half-integer spherical harmonics, for example: $Y_{1/2}^{1/2}(\theta, \varphi) \propto D_{0,\frac{1}{2}}^{\frac{1}{2}}(\varphi, \theta, \chi)$. This occurrence of internal projections different than $n = \sqrt{s(s+1)}$, which is the unique solution given for a two-dimensional angular momentum in Eq. (3), is due to the different possibilities of arranging two-dimensional angular momenta in the three-dimensional internal space of a particle. We show the three orthogonal arrangements of spin in the internal frame of a particle in Fig. 4, which are the only arrangements for spin-1/2, and the only stable arrangements for higher spin particles (i.e., all Standard Model particles are described by $n_{z'} = \sqrt{s(s+1)}$ or $n_{z'} = 0$). For the choice of quantization axis along $z'$, we have one state with $n_{z'} = \sqrt{s(s+1)}$ and doubly degenerate states with $n_{z'} = 0$. By symmetry, the same applies for choice of quantization axis along $x'$ or $y'$.

The three internal projections for spin-1/2 are the fewest states that form an isotropic set in three dimensions. These three states, if equally populated, form an isotropic combination in the internal coordinates, as shown by the spatial average:

$$\langle \hat{s}^2 \rangle = \frac{\sum_n \langle D_{n\,m}^s(\varphi,\theta,\chi) | 3\hat{s}_{z'}^2 | D_{n\,m}^s(\varphi,\theta,\chi) \rangle}{\sum_n \langle D_{n\,m}^s(\varphi,\theta,\chi) | D_{n\,m}^s(\varphi,\theta,\chi) \rangle} = \frac{\sum_n 3n^2}{3} = s(s+1) \tag{40}$$

The operation of $\hat{s}_{z'}^2$ on $D_{n\,m}^s(\varphi, \theta, \chi)$ returns the eigenvalue $n^2$, the $D_{n\,m}^s(\varphi, \theta, \chi)$ states are regularized with $\langle D_{n\,m}^s(\varphi,\theta,\chi) | D_{n\,m}^s(\varphi,\theta,\chi) \rangle = 1$ (see Section III), and $n$ is summed over the three states with values $n = 0$ (doubly degenerate) and $n = \sqrt{s(s+1)} = s$.

Another argument for the three projection states of $s = 1/2$ in Fig. 4 is to determine which set of $D_{n\,m}^{1/2}(\varphi, \theta, \chi)$ states are nonrotatable, as required. To achieve this, we consider the Wigner rotation matrix element for rotation of the $D_{n\,m}^{1/2}(\varphi, \theta, \chi)$ state to itself, given by $D_{n\,n}^{1/2}(\varphi, \theta, \chi)$. Although $D_{n\,n}^{1/2}(\varphi, \theta, \chi)$ is not normalizable for all $n$, we can determine expectation values of $D_{n\,n}^{1/2}(\varphi,\theta,\chi)$ using Eq. (22). For example, $\langle\cos\theta\rangle = n^2/s(s+1) = 4n^2/3$, so that for $n = \sqrt{3}/2 = s$, $\langle\cos\theta\rangle = 1$, which is consistent with $D_{s\,s}^{1/2}(\varphi,\theta,\chi) = \delta(\theta)$. This result gives all the probability at $\theta = 0$, so that the $D_{s\,m}^{1/2}(\varphi,\theta,\chi)$ state (with $s$ parallel to $z'$) is non-rotatable. In contrast, the states with $0 < n < s$ are not consistent with being nonrotatable, as $0 < \langle\cos\theta\rangle < 1$ for $D_{n\,n}^{1/2}(\varphi,\theta,\chi)$. However, for $n = 0$, the expectation value of all Legendre polynomials of $\cos\theta$ are zero for $k > 0$, $\langle P_k(\cos\theta)\rangle = 0$. Only the $k = 0$ term yields unity, which is consistent with $D_{0\,0}^{1/2}(\varphi,\theta,\chi) = 1$, so that the $D_{0\,m}^{1/2}(\varphi,\theta,\chi)$ state (with $s$ perpendicualar to $z'$) is also non-rotatable. Therefore, a consistent set of nonrotatable states are the $D_{n_{z'}\,m}^{1/2}(\varphi,\theta,\chi)$ for $n_{z'} = \sqrt{3}/2 = s$ and $n_{z'} = 0$. However, choosing $x'$ or $y'$ as the quantization axis yields the states $n_{x'} = \sqrt{3}/2 = s$ and $n_{x'} = 0$, and $n_{y'} = \sqrt{3}/2 = s$ and $n_{y'} = 0$, respectively. Therefore, we conclude that the three states shown in Fig. 4, parallel to each of the $x', y', z'$ axes, describe all these cases, with each $n = 0$ state describing the doubly degenerate states perpendicular to the quantization axis. Finally, we note that these three orthogonal states (where one is parallel to the symmetry axis of the particle, and the other two are perpendicular) are relativistically invariant, as a boost in the external frame will change the observed angle between two vectors only if the two vectors are *not* parallel or perpendicular to each other. Therefore, a boost will not transform the orthogonal states from one to another.

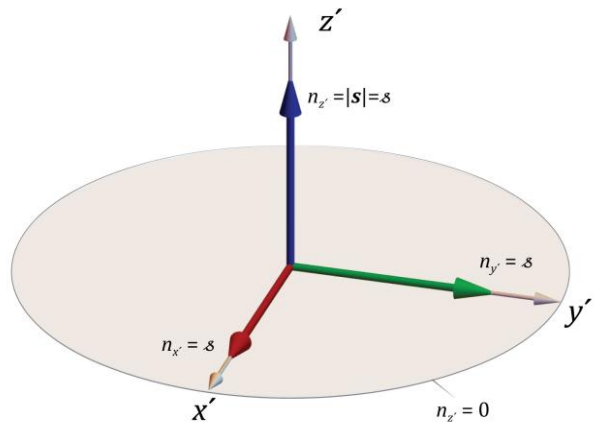

*Figure 4:* Depiction of the three orthogonal internal-coordinate spin-projection states, along the z axis with $n_{z'} = \sqrt{s(s+1)}$, the x axis with $n_{x'} = \sqrt{s(s+1)}$, and the y axis with $n_{y'} = \sqrt{s(s+1)}$; the latter two can be described by a doubly degenerate $n_{z'} = 0$ state.

The three internal projection states for $s = 1/2$ in Fig. 4 will be applied, in the next two sections, to spin, isospin, and weak isospin, to explain some of the Standard Model structure.

## VIII. IMPLICATIONS FOR STANDARD-MODEL PARTICLES: CLEBSCH-GORDAN COEFFICENTS IN THE INTERNAL FRAME

The charge conjugation operator, $\hat{C}$, reverses all the charges of a particle, and turns the wavefunction of a particle into that of its antiparticle. The interpretation given here (and explained further in Section X) is that operation of $\hat{C}$ is equivalent to the inversion of the internal coordinates of the particle: $\hat{P}_{in}(x',y',z') = (-x',-y',-z')$ causes a change in sign of the projection of $s$ on any of the $x', y', z'$ axes. An equivalent result is the inversion of the spin $s$, by $\hat{P}_{spin}(\chi) = -\chi$. Therefore, applying the spin-inversion operator $\hat{P}_{spin}$ to the spin wavefunction $D_{n\,m}^{s}(\varphi,\theta,\chi)$ of a particle yields:

$$\hat{P}_{spin}D_{n\,m}^{s}(\varphi,\theta,\chi) = D_{-n\,m}^{s}(\varphi,\theta,\chi) \qquad (41)$$

Thus, the wavefunction $D_{n\,m}^{s}(\varphi,\theta,\chi)$ is unchanged after operation of $\hat{P}_{spin}$ only for $n = 0$, and this a necessary condition for a particle to be its own antiparticle. Therefore, we propose that the Standard-Model particles fall into two categories, those with internal projection $n = s$, and with $n = 0$.

The Standard Model particles are presented in Table I according to their internal projection quantum number $n$. All particles, except the neutrino, clearly fall into one of the two categories: the charged particles have $n = s$, and the neutral particles that are their own antiparticle have $n = 0$. However, it has not been experimentally established whether the neutrino is its own antiparticle. We justify this categorization below, and show that the neutrino belongs to $n = s$. The justification will include calculations of Clebsch-Gordan coefficients in the internal frame.

**Table I:** The internal spin projection $n$ of Standard-Model particles, and proposed Dark-Matter particles

|  | Internal Spin projection $n$ | |
| --- | --- | --- |
|  | $n = \sqrt{s(s+1)} = s$ | $n = 0$ |
| Standard-Model particles | W boson<br>Quarks ($q$)<br>Charged Leptons ($e^-, \mu^-, \tau^-$)<br>Neutrinos ($\nu_{e^-}, \nu_{\mu^-}, \nu_{\tau^-}$) | Z boson<br>Photon ($\gamma$)<br>Gluon ($g$)<br>Higgs Boson ($H^0$) |
| Proposed Dark-Matter particles | Mirror Matter | Mirror Matter<br>Neutral Quarks ($q^0$)<br>Neutral Leptons ($e^0, \mu^0, \tau^0$) |

We begin by considering the emission of virtual photons, gluons, and Z bosons, in the reactions:

$$e^- \rightarrow e^- + \gamma \quad (42a)$$
$$q \rightarrow q + g \quad (42b)$$
$$e^- \rightarrow e^- + Z \quad (42c)$$

In all three cases, we can describe each process by the overlap integral of Eq. (21b), for which the fermion has the initial quantum numbers $s_i, n_i, m_i$ before emission, the final quantum numbers $s_f, n_f, m_f$ after emission, and the emitted boson has quantum numbers $s_b, n_b, m_b$:

$$\left\langle D^{s_i}_{n_i m_i}(R) \middle| D^{s_f}_{n_f m_f}(R) D^1_{n\, m_{f1}}(R) \right\rangle \propto \langle s_i\, n_i | s_f\, n_f, s_b\, n_b \rangle \langle s_i\, m_i | s_f\, m_f, s_b\, m_b \rangle \quad (43)$$

Conservation of the projection quantum numbers in the Clebsch-Gordan coefficients of Eq. (43) yields $m_i = m_f + m_b$ and $n_i = n_f + n_b$. As $n_b = 0$ (as expected for the particles $\gamma$, $g$, and $Z$ that are their own antiparticles), the $n$-state projection conservation is consistent with $n_i = n_f$. Note also that, for these cases, the Clebsch-Gordan coefficient $\langle s_i\, n_i | s_f\, n_f, s_b\, n_b \rangle = \left\langle \frac{1}{2}\frac{\sqrt{3}}{2} \middle| \frac{1}{2}\frac{\sqrt{3}}{2}, 1\, 0 \right\rangle = 1$ (which can be determined from Eq. 23b), leaving only the Clebsch-Gordan coefficient $\langle s_i\, m_i | s_f\, m_f, s_b\, m_b \rangle$ to determine the spatial aspects of the coupling process. This must be a general rule, that the magnitude $|\langle s_i\, n_i | s_f\, n_f, s_b\, n_b \rangle| = 1$ for all known allowed particle decays, otherwise we would introduce extra factors that would conflict with the experimental results of known particle physics. We will see that this rule holds, for all cases considered.

Next, we consider the decay of an $n = 0$ particle with spin $s_i = 1$, into a pair of $n = 0$ particles each with spin $s_f$. The decay probability will be proportional to the overlap integral of the spin wavefunctions:

$$\left\langle D^{s_i}_{0\,m_i}(R) | D^{s_f}_{0\,(m_i-m_f)}(R) D^{s_f}_{0\,m_f}(R) \right\rangle \propto \langle 1\,0 | s_f\,0, s_f\,0 \rangle \langle s_i\,m_i | s_f\,(m_i - m_f), s_f\,m_f \rangle \quad (44)$$

As the decay probability is proportional to the simple Clebsch-Gordan coefficient $\langle 1\,0 | s_f\,0, s_f\,0 \rangle$, we can draw some simple conclusions: $\langle 1\,0 | s_f\,0, s_f\,0 \rangle = 0$ for any value of $s_f$, because $\langle 1\,0 | s_f\,0, s_f\,0 \rangle \propto \langle s_f\,0 | 1\,0, s_f\,0 \rangle = 0$, using Eq. (23b). Therefore, the decay of an $n = 0$ particle with $s_i = 1$ *cannot* decay into two $n = 0$ particles. Therefore, Eq. (44) is a generalization of the Landau-Yang theorem [30,31,32], which is used to justify why, for example, the Z boson cannot decay to 2 photons (which are $n = 0$ particles). By using Eq. (44), we extend predictions beyond the Landau-Yang theorem, and predict that the Z boson cannot decay to any two $n = 0$ particles. As the Z boson is known to decay to neutrino-antineutrino pairs, it follows that the neutrino is *not* an $n = 0$ particle, and therefore that the neutrino is not a Majorana particle (which are their own antiparticles, and must have $n = 0$).

We next consider the decay of particles with spin $s_i$ and $n_i = \sqrt{s_i(s_i + 1)}$, to show that known decays are consistent with conservation of internal projection along the $z'$ axis of the decaying particle (and to determine the internal projection of the neutrino through another reaction). We consider the example of the decay of the $W^+$ boson:

$$W^+ \to e^+ + \nu_e \quad (45)$$

The application of Eq. (43) yields, for $s_W = 1$, $n_W = \sqrt{2}$:

$$\left\langle D^1_{\sqrt{2}\,m_W}(R) | D^{\frac{1}{2}}_{n'_{e^+},\,m_{e^+}}(R) D^{\frac{1}{2}}_{n'_{\nu_e}\,m_{\nu_e}}(R) \right\rangle \propto \left\langle 1\,\sqrt{2} \left| \tfrac{1}{2}\,n'_{e^+}, \tfrac{1}{2}\,n'_{\nu_e} \right.\right\rangle_{rms} \left\langle 1\,m_W \left| \tfrac{1}{2}\,m_{e^+}, \tfrac{1}{2}\,m_{\nu_e} \right.\right\rangle \quad (46)$$

where $n'_{e^+}$ and $n'_{\nu_e}$ are the projections of the spin of the positron and electron neutrino, respectively, onto the $z'$ axis of the decaying $W^+$ boson (and *not* on the symmetry axes of each product particle). Therefore, although conservation projection requires $n'_{e^+} + n'_{\nu_e} = n_W = \sqrt{2}$, the values of $n'_{e^+}$ and $n'_{\nu_e}$ must be integrated over their full allowed physics range. The calculation to determine $\left\langle 1\,\sqrt{2} \left| \tfrac{1}{2}\,n'_{e^+}, \tfrac{1}{2}\,n'_{\nu_e} \right.\right\rangle_{rms} = 1$, as expected for an allowed particle reaction, and is shown in Appendix C. The symmetry of the calculation determines that $n_{\nu_e} = n_{e^+} = \sqrt{3}/2$. A straightforward generalization of this calculation shows that the internal-frame Clebsch-Gordan coefficient, for the coupling of two angular momenta with $n_i = j_i$ to yield a third angular momentum with $n_3 = j_3$, is always unity. For example, for the coupling of orbital angular momenta $l_1 + l_2 = l_3$, the coefficent $\langle l_1\,n'_1, l_2\,n'_2 | l_3\,\ell_3 \rangle_{rms} = 1$, ensuring no change to known angular-momentum coupling rules.

We demonstrate $n_{\nu_e} = \sqrt{3}/2$ again, but in a more geometrically clear fashion. For the decay of a particle with $n_i = \sqrt{s_i(s_i + 1)}$, the projection of the spins along the decaying particle's $z'$ axis, and the subsequent projection on the space-fixed Z axis (given by $|s|\langle \cos\theta \rangle$) must be conserved for the reaction:

$$\frac{n_i m_i}{\sqrt{s_i(s_i + 1)}} = \sum_f \frac{n_f m_f}{\sqrt{s_f(s_f + 1)}} \quad (47)$$

where $\langle cos\theta \rangle = nm/[s(s+1)]$ from Eq. (22), and $|s| = \sqrt{s(s+1)}$. Note that here $n_f$ is the internal projection of each particle spin on its own $z'$ axis.

We use Eq. (47) for the decay of a $W^+$ boson in the $|SM\rangle = |11\rangle$ state, for which:

$$\frac{(n_{W^+})(m_{W^+})}{\sqrt{s_{W^+}(s_{W^+}+1)}} = \frac{(n_{e^+})(m_{e^+})}{\sqrt{s_{e^+}(s_{e^+}+1)}} + \frac{(n_{\nu_e})(m_{\nu_e})}{\sqrt{s_{\nu_e}(s_{\nu_e}+1)}} \tag{48a}$$

where $s_{W^+} = m_{W^+} = 1$, $s_{e^+} = m_{e^+} = \frac{1}{2}$, $s_{\nu_e} = m_{\nu_e} = \frac{1}{2}$, $n_{W^+} = \sqrt{s_{W^+}(s_{W^+}+1)} = \sqrt{2}$, $n_{e^+} = \sqrt{s_{e^+}(s_{e^+}+1)} = \sqrt{3}/2$, and note that $m_{W^+} = m_{e^+} + m_{\nu_e}$. Inserting these values into Eq. (48a) gives:

$$1 = \frac{1}{2} + \frac{1}{2}\left(\frac{n_{\nu_e}}{\sqrt{s_{\nu_e}(s_{\nu_e}+1)}}\right) \tag{48b}$$

and therefore that $n_{\nu_e} = \sqrt{s_{\nu_e}(s_{\nu_e}+1)} = \sqrt{3}/2$. Thus, we determine that the neutrino has the same internal projection as charged fermions, that the neutrino is a Dirac fermion, and its spin wavefunction is described by $D^{\frac{1}{2}}_{n\,m}(\varphi,\theta,\chi)$ with $n = \sqrt{3}/2$.

We next consider the decay of the Z boson to a fermion ($f$) anti-fermion ($\bar{f}$) pair, such as:

$$Z \to f + \bar{f} \tag{49}$$

The application of Eq. (43) yields, for $s_Z = 1, n_Z = 0$:

$$\left\langle D^1_{0\,m_W}(R) | D^{1/2}_{n'_f, m_f}(R) D^{1/2}_{n'_{\bar{f}}, m_{\bar{f}}}(R) \right\rangle \propto \left\langle 1\,0 \Big| \tfrac{1}{2}\,n'_f, \tfrac{1}{2}\,n'_{\bar{f}} \right\rangle \left\langle 1\,m_W \Big| \tfrac{1}{2}\,m_f, \tfrac{1}{2}\,m_{\bar{f}} \right\rangle \tag{50}$$

The calculation to determine $\left\langle 1\,0 \Big| \tfrac{1}{2}\,n'_f, \tfrac{1}{2}\,n'_{\bar{f}} \right\rangle_{rms} = 1$ is shown in Appendix C, as expected for an allowed particle reaction.

Finally, we consider the decay of a spin-0 particle, the Higgs boson $H^0$, into a fermion ($f$) anti-fermion ($\bar{f}$) pair, or boson anti-boson pairs, such as:

$$H^0 \to f + \bar{f} \tag{51a}$$
$$H^0 \to Z + Z \tag{51b}$$
$$H^0 \to \gamma + \gamma \tag{51c}$$
$$H^0 \to W^+ + W^- \tag{51d}$$

The Clebsch-Gordan coefficient for Eq. (51a) in the internal frame is given by $\left\langle 0\,0 \Big| \tfrac{1}{2}n_f, \tfrac{1}{2}n_{\bar{f}} \right\rangle$, where $n_f = -n_{\bar{f}}$. The value of $\langle j\,n, 0\,0 | j\,n \rangle$ is clearly 1, as is $\langle j\,m, 0\,0 | j\,m \rangle = 1$ in the external coordinates. However, exchanging the second and third angular momenta divides the coefficient by the square root of the degeneracy of the $|jm\rangle$ state, $\sqrt{2j+1}$, so that: $\langle j\,m, j-m | 0\,0 \rangle = (-1)^{j-m}/\sqrt{2j+1}$. However, the degeneracy of the $|jn\rangle$ state in the internal frame is 1. We find (and verify in Appendix C, for $s_2 = 0$

and $s_2 = 1$), that interchanging two spins does not change the magnitude of a Clebsch-Gordan coefficient in the internal frame, and can change it only by a phase factor:

$$|\langle s_1\, n_1, 0\, 0|s_3\, n_3\rangle| = |\langle s_1\, n_1, s_3\, n_3|0\, 0\rangle| = |\langle s_3\, n_3, 0\, 0|s_1\, n_1\rangle| \tag{52a}$$

$$|\langle s_1\, n_1, 1\, n_2|s_3\, n_3\rangle| = |\langle s_1\, n_1, s_3\, n_3|1\, n_2\rangle| = |\langle s_3\, n_3, 1\, n_2|s_1\, n_1\rangle| \tag{52b}$$

Therefore, $\left\langle 0\, 0\left|\frac{1}{2} n'_f, \frac{1}{2}\, n'_{\bar{f}}\right.\right\rangle_{rms} = 1$, within a phase factor, using Eq. (52a) and $\left\langle \frac{1}{2} n'_f\left|0\, 0, \frac{1}{2}\, n'_{\bar{f}}\right.\right\rangle = 1$, which is determined from Eq. (23b). An alternate derivation of this result, by integrating over all $n'_f$, is given in Appendix C.

The decays of Eqs (51b) and (51c) are described by $\langle 0\, 0|1 0, 1\, 0\rangle$. However, using Eq. (23b), $\langle 1\, 0|0\, 0, 1\, 0\rangle = 1$, and using Eq. (52) yields $|\langle 0\, 0|1 0, 1\, 0\rangle| = 1$. Finally, the decay of Eq. (51d) is described by $\langle 0\, 0|1 n'_b, 1\, n'_{\bar{b}}\rangle_{rms} = 1$, and this is also derived in Appendix C.

We note that all the Clebsch-Gordan coefficients in the internal frame are 1 (within a phase factor) for all the allowed particle reactions that we have considered, which is a necessary result to agree with known particle physics. In addition, the Clebsch-Gordan coefficient is 0 for reactions forbidden by the Landau-Yang theorem. The Clebsch-Gordan coefficients for the reactions discussed in this Section are summarized in Table II.

**Table II:** Internal spin frame Clebsch-Gordan coefficients for particle-decay reactions and orbital angular momentum coupling

| Decay Reaction or angular momentum coupling | Clebsch-Gordan coefficient in the internal frame $\langle s_i\, n_i|s_{f1}\, n_{f1}, s_{f2}\, n_{f2}\rangle$, within a phase factor |
|---|---|
| $e^- \to e^- + \gamma$ <br> $q \to q + g$ <br> $e^- \to e^- + Z$ | $\left\langle \frac{1}{2}\frac{\sqrt{3}}{2}\left|\frac{1}{2}\frac{\sqrt{3}}{2}, 1\, 0\right.\right\rangle = 1$ |
| $W^+ \to e^+ + \nu_e$ | $\left\langle 1\, \sqrt{2}\left|\frac{1}{2}\, n'_{e^+}, \frac{1}{2}\, n'_{\nu_e}\right.\right\rangle_{rms} = 1$ |
| $Z \to f + \bar{f}$ | $\left\langle 1\, 0\left|\frac{1}{2}\, n'_f, \frac{1}{2}\, n'_{\bar{f}}\right.\right\rangle_{rms} = 1$ |
| $H^0 \to f + \bar{f}$ | $\left\langle 0\, 0\left|\frac{1}{2}\, n'_f, \frac{1}{2}\, n'_{\bar{f}}\right.\right\rangle_{rms} = 1$ |
| $H^0 \to Z + Z$ <br> $H^0 \to \gamma + \gamma$ | $\langle 0\, 0|1\, 0, 1\, 0\rangle = 1$ |
| $H^0 \to W^+ + W^-$ | $\langle 0\, 0|1 n'_b, 1\, n'_{\bar{b}}\rangle_{rms} = 1$ |
| $Z \to \gamma + \gamma$ <br> $Z \to g + g$ | $\langle 1\, 0|1\, 0, 1\, 0\rangle = 0$ <br> (Forbidden) |
| $l_1 + l_2 \to l_3$ | $\langle l_1\, n'_1, l_2\, n'_2|l_3\, \ell_3\rangle_{rms} = 1$ |

In contrast, we note that the emission of bosons with spin $s > 1$ from a particle with angular momentum $j$ and $n = s$ will have a Clebsch-Gordan coefficient in the internal frame that is always greater than 1. For example, for emission of a spin-2 boson by a particle in the $|j\,n\rangle$ state, the Clebsch-Gordan coefficient in the internal frame is given by $\langle j\,n|2\,0,j\,n\rangle$, which is described by Eq. (23c). The value of $\langle j\,s|2\,0,j\,s\rangle$ is always greater than 1, because $P_2(cos\theta_n) = 1$ for $n = s$, however the coefficient is always greater than 1 (for finite $j \geq 1$):

$$\sqrt{\frac{j(j+1)}{(j-1/2)(j+3/2)}} = \sqrt{1 + \frac{3/4}{(j-1/2)(j+3/2)}} > 1 \tag{53}$$

For example, for $j = 1$, $\langle 1\,\sqrt{2}|2\,0,1\,\sqrt{2}\rangle = \sqrt{8/5}$ (see Appendix B2 for an explicit calculation of $\langle P_2(cos\theta)\rangle$ for the $D^1_{\frac{3}{2},1}(\varphi,\theta,\chi)$ state, which involves the coefficient $\langle 1\,3/2|2\,0,1\,3/2\rangle = \sqrt{19/10}$, which is also greater than 1). Similar arguments hold for bosons with higher spins, that the $\langle j\,s|s\,0,j\,s\rangle$ are greater than 1, for integer $s > 2$. In contrast, $\langle j\,0|s\,0,j\,0\rangle$ has a magnitude that is always less than 1. For example for $s = 2$ and $j = 1$, $\langle 1\,0|2\,0,1\,0\rangle = -\sqrt{2/5}$.

We saw that for Standard-Model particle reactions involving the bosons that are spin-1 (and for all Standard-Model particle reactions, shown in Table II), the internal-frame Clebsch-Gordan coefficients have a magnitude of unity, so that the internal projections $n$ seem to play the role of a hidden parameter that has no effect on known reactions. In contrast, the internal projections $n$ give internal-frame Clebsch-Gordan coefficients that differ from unity in reactions involving bosons with spins greater than 1. This may explain why all observed bosonic elementary particles have spin $s = 0$ or 1. At the very least, assuming the internal quantum number $n$ is correct, the internal-frame Clebsch-Gordan coefficients need to be taken into account for reactions with bosons with spin $s > 1$, such as for the spin-2 graviton. This may have implications for calculations in quantum gravity.

## IX. DARK MATTER CANDIDATES

Baryonic matter is associated with at least three obvious asymmetries: (1) the preponderance of matter over antimatter; (2) the fact that known baryonic matter violates parity through the weak interaction; and (3) the further asymmetry introduced here, that known baryons have internal angular momentum projection $n = \sqrt{s(s+1)}$ only, which constitutes a lack of isotropy in the internal (body-fixed) coordinate system (due to lack of baryons with $n = 0$, needed for isotropy in the internal coordinate system).

Mirror Matter [33] has been proposed as a solution to asymmetry (2), which results in a doubling of all Standard Model particles with mirror pairs, that are dark with respect to normal baryonic matter, and have opposite parity violation effects. However, Mirror Matter particles will be as self-interacting as normal baryonic matter is, and so can only constitute a small fraction of the dark matter (assuming equal temperature with baryonic matter), as the $\Lambda$CDM model assumes non-interacting dark matter, and it explains the cosmic background radiation power spectrum well [34] (we note that Mirror Matter can explain dark matter, for low temperatures of the Mirror Matter [35]). Nevertheless, there is evidence that suggests the existence of some self-interacting dark matter [36,37,38,39,40], and Mirror Matter offers a plausible explanation for such self-interacting dark matter.

We address the lack of internal projection isotropy, i.e. asymmetry (3), by proposing that there exists neutral baryonic matter, based on neutral quarks that have internal projection $n = 0$, and with similar mass as charged baryonic matter. We also address the parity asymmetry (2) by proposing equal amounts of mirror matter. Finally, we can also address the matter-antimatter asymmetry (1) by proposing that the mirror matter is also antimatter (related to baryonic matter by a $\hat{C}\hat{P}$ transformation); although there currently seems to be no predictive power in the proposal of mirror antimatter, we mention it as a plausible way to resolve the matter-antimatter asymmetry. In this way, all three asymmetries are addressed, albeit asymmetries (1) and (2) are entangled; however, it seems that this is the highest symmetry for which the matter and antimatter do not annihilate.

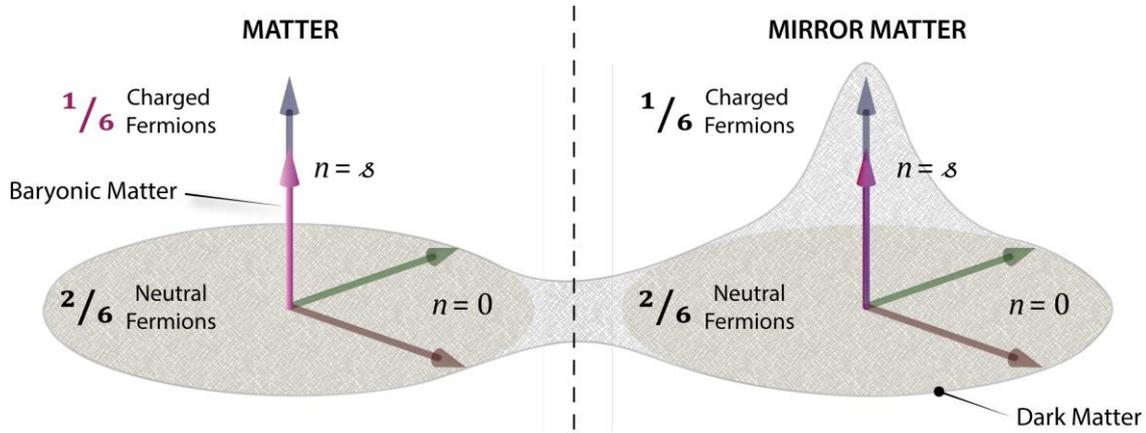

*Figure 5:* The proposed abundance of particles: 1/6 is baryonic matter, and the remaining 5/6 is dark matter (shaded gray). Of the 5/6 dark matter, 1/6 is mirror matter, 2/6 neutral dark matter, and 2/6 mirror neutral dark matter; therefore, 4/6 of the dark matter is cold dark matter (CDM), and 1/6 (the baryonic mirror matter) is self-interacting.

The abundance of particles of the proposed symmetries is shown in Fig. 5. We use two assumptions to explain the observed fraction of dark matter: (1) there are equal amounts of matter and mirror matter, and (2) the populations of the internal-frame $n$ states are isotropic, with equal population. Therefore, there are 6 equal parts, shown by the 2 pairs of 3 orthogonal arrows in Fig. 5, where only 1/6 corresponds to baryonic matter, and the remaining 5/6 are dark matter. Specifically, 1/6 is baryonic matter (the pink arrow at left with $n = s$), 1/6 is mirror matter (the pink arrow at right with $n = s$), 2/6 are baryonic dark matter (the two green and brown arrows with $n = 0$ at left), and 2/6 of mirror dark matter (the two green and brown arrows with $n = 0$ at right). Note that the mirror matter is dark to the baryonic matter. Therefore, the predicted fraction that is dark matter is $5/6 \approx 83.3\%$, which agrees well with experimental observations of 85% [34]. We note that if the mass of the ($n = 0$) neutral quarks are larger than the charged quarks, by a similar ratio as that of the ($n = 0$) Z boson to the ($n = \sqrt{S(S+1)}$) W boson (by about 13%), then the calculated dark matter fraction increases from 5/6 to about 85%, to give near exact agreement, shown by the following ratio: $(1 + 4 \times 1.13)/(2 + 4 \times 1.13) \approx 85\%$.

We now consider whether these neutral ($n = 0$) baryons can be observed directly, either through direct decay, or produced at particle colliders. First, they cannot decay to any known Standard-Model particles, because it is not possible to do so while conserving charge, lepton number, and baryon number. Second, we showed, using Eq. (44), that the decay of a Z boson is forbidden to any $n = 0$ particle-antiparticle pair (such as two photons or two neutral quarks); this explains why neutral ($n = 0$) quarks cannot be formed from Z boson decay. Third, a channel for potentially observing evidence for these dark-matter particles is the formation of neutral quark-antiquark pairs from the decay of the Higgs boson. If the neutral quarks have similar masses with the charged quarks, the Higgs neutral-quark decay channel will be dominated by neutral bottom quarks. Current searches for dark channels limit their fraction to about 18% (2σ = 10%) [41]. A calculation of the expected production of neutral quark-antiquark pairs can be compared to this dark-channel uncertainty, to test this possibility. Furthermore, future experiments will reduce the dark-channel uncertainty further, and may confirm or disprove this prediction.

An additional prediction that can be tested, is that the self-interacting mirror antimatter is 1/6 of the total matter (from Fig. 5, equal to the baryonic contribution). It can be investigated whether the tension in the current dark-matter models is eliminated or reduced from the inclusion of the self-interacting mirror antimatter (as 1/5 of the total dark matter).

## X. INTERNAL PROJECTIONS FOR THE STRONG AND WEAK FORCES

In this paper, we have proposed that the spin of fundamental particles have three internal projections, described by $n_{x'} = n_{y'} = n_{z'} = s$ (or equivalently, $n_{z'} = s$ and the doubly degenerate $n_{z'} = 0$). Spin is associated with the electromagnetic force; there are very close analogues of spin for the strong and weak forces, isospin and weak isospin, respectively, and it is natural to consider whether the three internal projections also exist for the spins of these forces. We propose that they do, and that they are already clearly present in the Standard Model: the three internal projections of isospin are the three quark colors, and the three internal projections of weak isospin correspond to the three generations of matter. We present the similar projection structure for the three forces in Fig. 6, to highlight the similarities and the differences.

In all three cases, we propose (without theoretical justification, but as a simple explanation of the observations) that there is a charge axis that breaks the symmetry between the three projection states, and that, the projection of the spin on the charge axis determines the charge of each state. For the electromagnetic force, one projection of the spin is fully parallel to the electric charge axis, so that the particles of this state have maximum electric charge $Q$ for the particle in question (given by $Q = T_3 + \frac{1}{2}Y_W$, where $T_3$ is the third component of the weak isospin and $Y_W$ is the weak hypercharge). In contrast, the other two orthogonal states have 0 projection on the charge axis, and thus are neutral. These neutral spin-1/2 particles (e.g. quarks) correspond to dark matter (see section IX and Fig. 5).

For the strong force, each of the three spin projections correspond to the three colors of quarks. The (1,1,1) vector, which corresponds to the sum of all three projection vectors, is perpendicular to the strong-charge axis. The arrangement of the three states is such that each state has an fractional projection on the charge field, and thus has a fractional strong charge, which is forbidden for each free particle, as it must be quantized (have integer value). However, either a quark-antiquark pair or three quarks of each projection (color) yields zero projection, and hence a zero net strong charge, which is correctly quantized and allowed. Therefore, Fig. 6b gives a geometric interpretation of the known rules for quark combinations, to yield a zero projection (colorless) combination for allowed composite particles. We can also understand the gluon in terms of these internal projections. We consider coupling a spin-1/2 particle with a spin-1/2 antiparticle, which will give a spin-1 and a spin-0 particle: $\frac{1}{2} \otimes \frac{1}{2} = 0 \oplus 1$, corresponding to m-state projections of a singlet and triplet: $2 \otimes 2 = 1 \oplus 3$. However, the three internal $n$-state projections of the particle and the three internal n-state projections of the antiparticle will give nine total projections, of which will correspond to an isospin-0 particle (singlet) and an octet for the isospin-1 particle: $3 \otimes 3 = 1 \oplus 8$. Therefore, the remaining eight states correspond to the gluon (in this case), showing consistency with the known SU(3) symmetry of the strong force.

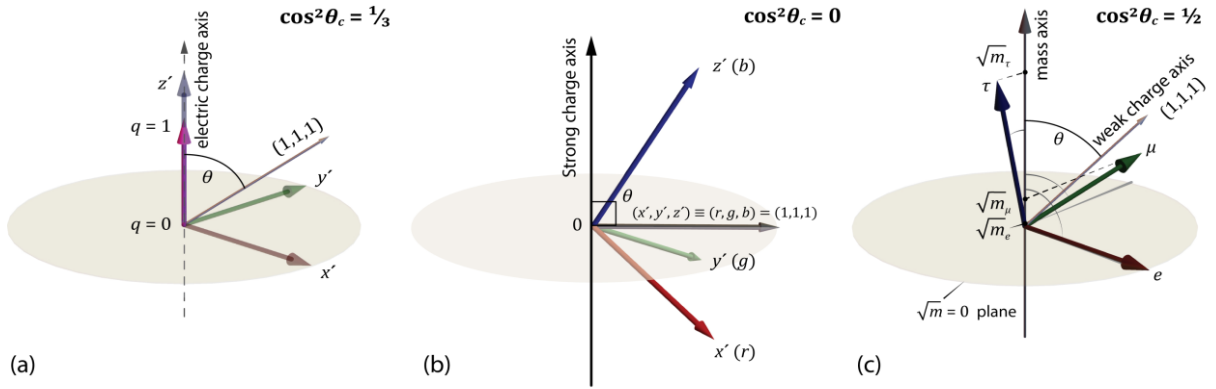

*Figure 6:* The three spin-projection states for (a) spin for the electromagnetic force, (b) isospin for the strong force, and (c) weak isospin for the weak force. See text for details.

For the weak force, we propose that the internal weak isospin projections play a role in determining the particle mass of the fermions, and thus these three projections produce the three generations of matter (for the quarks and the leptons). For the charged leptons, these three projections correspond to the electron, muon, and tau. Koide noticed in 1981 that the masses of the charged leptons are related by a simple formula [42]:

$$\frac{m_e + m_\mu + m_\tau}{\left(\sqrt{m_e} + \sqrt{m_\mu} + \sqrt{m_\tau}\right)^2} \approx \frac{2}{3} \qquad (54)$$

Experimental measurements of the charged lepton masses agree with Eq. (54) within experimental uncertainty. A geometric interpretation of Eq. (54) was proposed by Foot [43], that the (1,1,1) vector is at an angle of 45° to the (square-root) mass axis. This interpretation, shown in Fig. 6c, is consistent with the space of the three internal spin projections given here. This formula, or extensions of it, have also been applied to quarks and neutrinos [44,45,46]. For example, the three heaviest quarks are found to satisfy Eq. (54) [44,46]. In addition, we propose a weak charge axis parallel to the (1,1,1) vector, so that the charged leptons have the same weak charge.

Note that for the description in Fig. 6, the charge conjugation operator, $\hat{C}$, is equivalent to inverting all the charge axes, or inverting all the spin projections, as in both cases this inverts all the charges of a particle. Therefore, $\hat{C}$ is equivalent to the internal-frame parity inversion operator: $\hat{C}(x',y',z') = (-x',-y',-z')$.

In general, we see that the internal projections seem to play a similar role in all three forces, in determining the charge or mass of particles, but the character of the symmetry breaking is different for each of them, with the square of the cosine of the angle between the (1,1,1) vector and the charge/mass axis (shown in Fig. 6a, 6b, and 6c) being 1/3, 0, and 1/2, respectively. In our opinion, this similarity makes the argument for the existence of dark matter more compelling, as its absence would represent a symmetry breaking of electromagnetism with respect to the other two forces. It is also remarkable to notice that the internal spin projections of the three forces generates much of the observed structure of the Standard Model. However, a comprehensive theory of particle physics is needed which will explain many missing details, such as the angle $\theta_c$ for each force, shown in Figs. 6a, 6b, and 6c.

## XI. CONCLUSIONS

We have presented the spatial wavefunctions of spin, $D^s_{n\,m}(\varphi,\theta,\chi)$, that treat $s$ as a three-dimensional entity in the external frame, but as a two-dimensional entity in the internal frame. Compared to the conventional $|sm\rangle$ spin states of elementary particles, these wavefunctions have an additional three internal body-fixed spin projections: $n_{x'} = n_{y'} = n_{z'} = \sqrt{s(s+1)}$ (or equivalently, $n_{z'} = \sqrt{s(s+1)}$, and the doubly degenerate $n_{z'} = 0$ state).

These spatial spin wavefunctions $D^s_{n\,m}(\varphi,\theta,\chi)$ reproduce all the known results of quantum mechanical angular momentum theory. However, the additional degree of freedom, the internal projection $n$, allows the explanation of new phenomena, or the explanation of known phenomena geometrically. We summarize five advantages of the $D^s_{n\,m}(\varphi,\theta,\chi)$ angular momentum wavefunctions:

(1) The $D^l_{n\,m}(\varphi,\theta,\chi)$, for $n = \sqrt{l(l+1)}$, connect quantum mechanical and classical angular momentum smoothly, through the asymptotic wavefunction of Eq. (3), described in Ref. [9].

(2) The gyromagnetic ratio of charged particles with spin $s$, $g = 2$, can be calculated as a direct expectation value of the magnetic-moment operator for the $D^s_{n\,m}(\varphi,\theta,\chi)$ states, with $n = \sqrt{s(s+1)} = s$.

(3) The three $n$-state projections, $n_{x'} = n_{y'} = n_{z'} = s$, for Standard-Model particles suggest an

explanation for some of the Standard-Model structure. For each of the three forces of nature (leaving gravity aside), the three internal spin projections correspond to: (a) the three quark colors, for isospin of the strong force; (b) the three generations of matter for weak isospin of the weak force; (c) charged matter with $n_z = s$ and doubly-degenerate neutral dark matter (with $n_z = 0$), for spin of the electromagnetic force. Furthermore, all Standard-Model particles can be categorized as either $n_{z'} = s$ or $n_{z'} = 0$.

(4) The internal projection $n$ plays a role in the selection rules of decaying elementary particles, and can generalize the Landau-Yang theorem [30,31]. Using the selection rule of Eq. (44), or the internal-projection conservation of Eq. (46), we show that the neutrino is described by $n = s$ (for both spin and weak isospin), and is thus a Dirac fermion, and not a Majorana fermion (which has $n = 0$).

(5) The new degree of freedom of the internal projection $n$ allows new candidates for dark matter, such as neutral quarks (with $n = 0$). Dark matter fractions of about 5/6 can be predicted, assuming neutral quarks have a similar mass as charged quarks, and assuming the existence of equal quantities of mirror matter.

Some of these points allow us to make testable predictions or hypotheses that go beyond standard angular-momentum theory, e.g, the Dirac-fermion nature of the neutrino, some decay selection rules of fundamental particles, and the prediction of properties of new dark-matter candidates and their self-interaction in cosmological models. Most importantly, these predictions are expected to be tested in the next few years, as there are intensive experimental efforts related to all three predictions.

We believe that the importance of the potential applications of some or all of the five points above make it worthwhile to study the foundations of the spatial wavefunctions of spin $D^s_{n\,m}(\varphi,\theta,\chi)$ further, and to investigate whether the $D^s_{n\,m}(\varphi,\theta,\chi)$ are fully consistent with quantum field theory and the Standard Model.

## ACKNOWLEDGEMENTS

This work was partially supported by the Hellenic Foundation for Research and Innovation (HFRI) and the General Secretariat for Research and Technology (GSRT), under grant agreement No. HFRI-FM17-3709 (project NUPOL). We thank Dr. Giorgos Katsoprinakis for assistance with the figures, Vasilis Niarchos for constructive criticism, and Vasiliki Pavlidou, Kostas Tassis, and George Kastrinakis for making helpful suggestions.

## APPENDIX A

Here we show that, for all $n$:

$$\hat{J}_-\hat{J}_+ D^j_{n,j}(\varphi,\theta,\chi) = 0 \tag{A1}$$

$D^j_{n,j}(\varphi,\theta,\chi)$ is defined in Eq. (5), and $d^j_{n,j}(\theta)$ is given in Eq. (10), where there are two terms, proportional to the constants $A$ and $B$. For the first term, for $m = j$, ${}_2F_1\left(m-j, m+j+1; 1+m-n; \sin^2\frac{\theta}{2}\right) = 1$, so that the first term reduces to $Ae^{ij\varphi}\left(\sin\frac{\theta}{2}\right)^{j-n}\left(\cos\frac{\theta}{2}\right)^{j+n} e^{in\chi}$. Operation of $\hat{J}_+$ on this term, from Eq. (12), gives 0, satisfying Eq. (A1) for all $n$. Operation of $\hat{J}_+$ on the second term yields:

$$\hat{J}_+ B\left(\sin\frac{\theta}{2}\right)^{n-j}\left(\cos\frac{\theta}{2}\right)^{n+j} {}_2F_1\left(n-j, n+j+1; 1+n-j; \sin^2\frac{\theta}{2}\right)$$

$$= Be^{i(j+1)\varphi}(n-j)\left(\sin\frac{\theta}{2}\right)^{-j+n-1}\left(\cos\frac{\theta}{2}\right)^{-j-n-1} e^{in\chi} \tag{A2}$$

Finally, the operation of $\hat{J}_-$ on the result of Eq. (A2) yields 0, showing that Eq. (A1) holds for all $n$, and by symmetry, $\hat{J}_+\hat{J}_- D^j_{n,-j}(\varphi,\theta,\chi) = 0$.

## APPENDIX B

We demonstrate the use of Eqs. (19) and (22), by calculating the expectation values of $\cos\theta$ for $|\Psi\rangle = D^{\frac{1}{2}}_{1,\frac{1}{2}}(\varphi,\theta,\chi)$, and of $P_2(\cos\theta)$ for $|\Psi\rangle = D^1_{\frac{3}{2},1}(\varphi,\theta,\chi)$. We will calculate the expectation values separately for each of the two terms of the wavefunction, given in Eqs. (17) and (18), to show that both terms of each wavefunction give the same results for both normalization and the expectation values, and are, therefore, practically symmetric.

### B1. The $D^{\frac{1}{2}}_{1,\frac{1}{2}}(\varphi,\theta,\chi)$ state

First, we calculate $\Psi^*\Psi \sin\theta$, from Eq. (17), for the integral of Eq. (19a):

$$\Psi^*\Psi \sin\theta = |N_{1/2}|^2 \left[2\left(\cos\frac{\theta}{2}\right)^4 + \frac{1}{2}\left(\cos\frac{\theta}{2}\right)^{-2}\left(\sin\frac{\theta}{2}\right)^2 (2+\cos\theta)^2\right] \tag{B1}$$

Then, we expand the second term (which diverges at $\theta = \pi$), in powers of $\sin\theta$:

$$\frac{1}{2}\left(\cos\frac{\theta}{2}\right)^{-2}\left(\sin\frac{\theta}{2}\right)^2 (2+\cos\theta)^2 = \frac{\left(\sin\frac{\theta}{2}\right)^4 (2+\cos\theta)^2}{2\left(\sin\frac{\theta}{2}\right)^2\left(\cos\frac{\theta}{2}\right)^2} = \frac{(1-\cos\theta)^2(2+\cos\theta)^2}{2\sin^2\theta}$$

$$= \frac{1-\cos\theta}{\sin^2\theta} + \frac{1 - 2\cos\theta + \sin^2\theta}{2} \tag{B2}$$

Using Eq. (19b), we subtract off the diverging first term, where $f(\theta) = (1-\cos\theta)/\sin^2\theta$.

Finally, we perform the integral of Eq. (19a) separately on the first term and regularized second term to

yield:

$$\int_0^\pi d\theta \left[\int_0^{2\pi} d\varphi \int_0^{2\pi q} \lim_{q\to\infty} \frac{1}{q} d\chi\, 2\left|N_{\frac{1}{2}}\right|^2 \left(\cos\frac{\theta}{2}\right)^4\right] = 3\pi^3 |N_{1/2}|^2 \tag{B3a}$$

$$\int_0^\pi d\theta \left[\int_0^{2\pi} d\varphi \int_0^{2\pi q} \lim_{q\to\infty} \frac{1}{q} d\chi\, \left|N_{\frac{1}{2}}\right|^2 \left(\frac{1 - 2\cos\theta + \sin^2\theta}{2}\right)\right] = 3\pi^3 |N_{1/2}|^2 \tag{B3b}$$

Showing that both terms of $D^{\frac{1}{2}}_{1,\frac{1}{2}}(\varphi,\theta,\chi)$ have the same norm, and that the sum of both is $6\pi^3 |N_{1/2}|^2$.

We now follow a similar procedure for the calculation of $\langle \Psi | \cos\theta | \Psi \rangle$:

$$\Psi^* \cos\theta\, \Psi \sin\theta = |N_{1/2}|^2 \cos\theta \left[2\left(\cos\frac{\theta}{2}\right)^4 + \frac{1}{2}\left(\cos\frac{\theta}{2}\right)^{-2}\left(\sin\frac{\theta}{2}\right)^2 (2+\cos\theta)^2\right] \tag{B4}$$

Again, we expand the second term (which diverges at $\theta = \pi$), in powers of $\sin\theta$:

$$\frac{\cos\theta}{2}\left(\cos\frac{\theta}{2}\right)^{-2}\left(\sin\frac{\theta}{2}\right)^2 (2+\cos\theta)^2 = \frac{\cos\theta(1-\cos\theta)^2(2+\cos\theta)^2}{2\sin^2\theta}$$

$$= \frac{-1+\cos\theta}{\sin^2\theta} + \frac{\cos\theta + 2\sin^2\theta + \cos\theta \sin^2\theta}{2} \tag{B5}$$

We subtract off the diverging first term, where $f(\theta) = -(1 - \cos\theta)/\sin^2\theta$.

Finally, we perform the integral of Eq. (19a) separately on the first term and regularized second term to yield:

$$\int_0^\pi d\theta \left[\int_0^{2\pi} d\varphi \int_0^{2\pi q} \lim_{q\to\infty} \frac{1}{q} d\chi\, \left|N_{\frac{1}{2}}\right|^2 2\cos\theta \left(\cos\frac{\theta}{2}\right)^4\right] = 2\pi^3 |N_{1/2}|^2 \tag{B6a}$$

$$\int_0^\pi d\theta \left[\int_0^{2\pi} d\varphi \int_0^{2\pi q} \lim_{q\to\infty} \frac{1}{q} d\chi\, \left|N_{\frac{1}{2}}\right|^2 \left(\frac{\cos\theta + 2\sin^2\theta + \cos\theta \sin^2\theta}{2}\right)\right] = 2\pi^3 \left|N_{\frac{1}{2}}\right|^2 \tag{B6b}$$

The expectation values of both terms yield the same results in Eq. (B6), and their sum is $4\pi^3 \left|N_{\frac{1}{2}}\right|^2$.

Finally, the expectation value $\langle \cos\theta \rangle$ for $|\Psi\rangle = D^{\frac{1}{2}}_{1,\frac{1}{2}}(\theta)$ is given by the ratio of the results of the integrals of Eqs. (B6) and (B3), which is $\langle \cos\theta \rangle = 2/3$. This agrees with $\langle \cos\theta \rangle = nm/s(s+1)$, the analytical result from Eqs. (22) and (23b), for $s = m = 1/2$ and $n = 1$.

## B2. The $D^1_{\frac{3}{2},1}(\varphi,\theta,\chi)$ state

First, we calculate $\Psi^*\Psi \sin\theta$, from Eq. (18), for the integral of Eq. (19a):

$$\Psi^*\Psi \sin\theta = |N_1|^2 \left[ 2\left(\cos\frac{\theta}{2}\right)^6 + \frac{1}{2}\left(\cos\frac{\theta}{2}\right)^{-4}\left(\sin\frac{\theta}{2}\right)^2\left(\frac{9}{4}+\frac{3}{2}\cos\theta - \frac{1}{2}\sin^2\theta\right)^2 \right] \quad (B7)$$

Then, we expand the second term (which diverges at $\theta = \pi$), in powers of $\sin\theta$:

$$\frac{1}{2}\left(\cos\frac{\theta}{2}\right)^{-4}\left(\sin\frac{\theta}{2}\right)^2\left(\frac{9}{4}+\frac{3}{2}\cos\theta - \frac{1}{2}\sin^2\theta\right)^2 = \frac{(1-\cos\theta)^3\left(\frac{9}{4}+\frac{3}{2}\cos\theta-\frac{1}{2}\sin^2\theta\right)^2}{\sin^4\theta}$$

$$= \frac{36 - 36\cos\theta - 3\sin^2\theta - 15\cos\theta\sin^2\theta}{16\sin^4\theta} + \frac{1 - 4\cos\theta + 3 + \cos\theta\sin^2\theta}{4} \quad (B8)$$

Using Eq. (19b), we subtract off the diverging first term, inversely proportional to $\sin^4\theta$, where $f(\theta) = (36 - 36\cos\theta - 3\sin^2\theta - 15\cos\theta\sin^2\theta)/16\sin^4\theta$.

Finally, we perform the integral of Eq. (19a) separately on the first term and regularized second term to yield:

$$\int_0^\pi d\theta \left[\int_0^{2\pi} d\varphi \int_0^{2\pi q} \lim_{q\to\infty}\frac{1}{q}d\chi\, 2|N_1|^2\left(\cos\frac{\theta}{2}\right)^6\right] = \frac{5}{2}\pi^3 |N_{1/2}|^2 \quad (B9a)$$

$$\int_0^\pi d\theta \left[\int_0^{2\pi} d\varphi \int_0^{2\pi q} \lim_{q\to\infty}\frac{1}{q}d\chi\, |N_1|^2\left(\frac{1 - 4\cos\theta + 3 + \cos\theta\sin^2\theta}{4}\right)\right] = \frac{5}{2}\pi^3 |N_{1/2}|^2 \quad (B9b)$$

Showing that both terms of $D^{\frac{1}{2}}_{1,\frac{1}{2}}(\varphi,\theta,\chi)$ have the same norm, and that the sum of both is $5\pi^3|N_{1/2}|^2$.

We now follow a similar procedure for the calculation of $\langle\Psi|P_2(\cos\theta)|\Psi\rangle$:

$$\Psi^* P_2(\cos\theta)\Psi \sin\theta = P_2(\cos\theta)|N_1|^2 \left[ 2\left(\cos\frac{\theta}{2}\right)^6 + \frac{\left(\sin\frac{\theta}{2}\right)^2\left(\frac{9}{4}+\frac{3}{2}\cos\theta-\frac{1}{2}\sin^2\theta\right)^2}{2\left(\cos\frac{\theta}{2}\right)^4} \right] \quad (B10)$$

Again, we expand the second term (which diverges at $\theta = \pi$), in powers of $\sin\theta$:

$$\frac{P_2(\cos\theta)\left(\sin\frac{\theta}{2}\right)^2\left(\frac{9}{4}+\frac{3}{2}\cos\theta-\frac{1}{2}\sin^2\theta\right)^2}{2\left(\cos\frac{\theta}{2}\right)^4} = \frac{P_2(\cos\theta)(1-\cos\theta)^3\left(\frac{9}{4}+\frac{3}{2}\cos\theta-\frac{1}{2}\sin^2\theta\right)^2}{\sin^4\theta}$$

$$= \frac{36-36\cos\theta-57\sin^2\theta+39\cos\theta\sin^2\theta}{16\sin^4\theta} + \frac{17+13\cos\theta+12\sin^2\theta+56\cos\theta\sin^2\theta-36\sin^4\theta-12\cos\theta\sin^4\theta}{32} \quad (B11)$$

We subtract off $f(\theta) = (36 - 36\cos\theta - 57\sin^2\theta + 39\cos\theta\sin^2\theta)/16\sin^4\theta$ from the diverging first term.

Finally, we perform the integral of Eq. (19a) separately on the first term and regularized second term to yield:

$$\int_0^\pi d\theta \left[4\pi^2 |N_1|^2 \, 2P_2(\cos\theta)\left(\cos\frac{\theta}{2}\right)^6\right] = \frac{19}{16}\pi^3 |N_{1/2}|^2 \quad (B12a)$$

$$\int_0^\pi d\theta \left[ 4\pi^2 |N_1|^2 \left( \frac{17 + 13\cos\theta + 12\sin^2\theta + 56\cos\theta\sin^2\theta - 36\sin^4\theta - 12\cos\theta\sin^4\theta}{32} \right) \right]$$

$$= \frac{19}{16}\pi^3 |N_1|^2 \tag{B12b}$$

The expectation values of both terms yield the same results in Eq. (B12), and their sum is $\frac{19}{8}\pi^3 |N_1|^2$.

Finally, the expectation value $\langle P_2(\cos\theta) \rangle$ for $|\Psi\rangle = D^1_{\frac{3}{2},1}(\varphi,\theta,\chi)$ is given by the ratio of the results of the integrals of Eqs. (B12) and (B9), which is $\langle P_2(\cos\theta) \rangle = 19/40$. This agrees with the analytical result from Eqs. (22) and (23c), for $s = m = 1$ and $n = 3/2$:

$$\langle P_2(\cos\theta) \rangle = \frac{s(s+1)}{\left(s - \frac{1}{2}\right)\left(s + \frac{3}{2}\right)} \left( \frac{3m^2 - s(s+1)}{2s(s+1)} \right) \left( \frac{3n^2 - s(s+1)}{2s(s+1)} \right) \tag{B13}$$

**APPENDIX C**

In this Appendix, we calculate the internal-frame Clebsch-Gordan coefficients for $\langle 1\ \sqrt{2} | \frac{1}{2}\ n'_{e^+}, \frac{1}{2}\ n'_{\nu_e} \rangle$, $\langle 1\ 0 | \frac{1}{2}\ n'_f, \frac{1}{2}\ n'_{\bar{f}} \rangle$, $\langle 0\ 0 | \frac{1}{2}\ n'_f, \frac{1}{2}\ n'_{\bar{f}} \rangle$, and $\langle 0\ 0 | 1\ n'_f, 1\ n'_{\bar{f}} \rangle$ in Table II and Eqs. (46) and (50). We use analytical expressions for the squares of the usual Clebsch-Gordan coefficient in the external frame, as the derivation of Clebsch-Gordan coefficients can be performed independent of the use of raising and lowering operators [47,48], and therefore applies to the coupling of angular momenta in the internal frame.

The Clebsch-Gordan coefficients in the external frame are given by a general, analytical expression, first given by Racah [18]. However, if we choose a fixed value of $j_3$, then this complicated, general expression can be simplified, particularly if $j_3$ is close to its maximal value. First, we use the analytical expression for the square of the Clebsch-Gordan coefficient $\langle j_1 m_1, j_2 m_2 | j_3 m_3 \rangle$ with the maximal value $j_3 = j_1 + j_2$ [49]:

$$\langle j_1 m_1, j_2 m_2 | j_1 + j_2, m_1 + m_2 \rangle^2 = \frac{(2j_1)!(2j_2)!(j_1 + j_2 + m_1 + m_2)!(j_1 + j_2 - m_1 - m_2)!}{(2j_1 + 2j_2)!(j_1 + m_1)!(j_1 - m_1)!(j_2 + m_2)!(j_2 - m_2)!} \tag{C1}$$

For the coefficient $\langle 1\ \sqrt{2} | \frac{1}{2}\ n'_{e^+}, \frac{1}{2}\ n'_{\nu_e} \rangle$, projection conservation gives $n'_{e^+} + n'_{\nu_e} = \sqrt{2}$. Setting $x = n'_{e^+}$, we have $n'_{\nu_e} = \sqrt{2} - x$. Therefore, $\langle 1\ \sqrt{2} | \frac{1}{2}\ n'_{e^+}, \frac{1}{2}\ n'_{\nu_e} \rangle^2$ can be determined as an integral over the regularizable projections in the internal frame, for the range $-\infty < x < \infty$ (note that Eq. (19) ensures regularizability over this range):

$$\langle 1\ \sqrt{2} | \frac{1}{2}\ n'_{e^+}, \frac{1}{2}\ n'_{\nu_e} \rangle^2 = \int_{-\infty}^{\infty} \langle \frac{1}{2}\ x, \frac{1}{2}\ \sqrt{2} - x | 1\ \sqrt{2} \rangle^2 dx \tag{C2}$$

where $\left\langle \frac{1}{2} x, \frac{1}{2} \sqrt{2} - x \middle| 1 \sqrt{2} \right\rangle^2$ is determined from Eq. (C1):

$$\left\langle \frac{1}{2} x, \frac{1}{2} \sqrt{2} - x \middle| 1 \sqrt{2} \right\rangle^2 = \frac{(1-\sqrt{2})!(1+\sqrt{2})!}{2(\frac{1}{2}+x)!(\frac{1}{2}-x)!(\frac{1}{2}-\sqrt{2}+x)!(\frac{1}{2}+\sqrt{2}-x)!} \quad (C3)$$

Integrating Eq. (C2) can be performed using (C3) and the integral [50]:

$$\int_{-\infty}^{\infty} \frac{dx}{\Gamma(\alpha+x)\Gamma(\beta-x)\Gamma(\gamma+x)\Gamma(\delta-x)} = \frac{\Gamma(\alpha+\beta+\gamma+\delta-3)}{\Gamma(\alpha+\beta-1)\Gamma(\beta+\gamma-1)\Gamma(\gamma+\delta-1)\Gamma(\delta+1-1)} \quad (C4)$$

where $\Gamma(n) = (n-1)!$, and $Re[\alpha + \beta + \gamma + \delta] > 3$. The integral yields $\left\langle 1 \sqrt{2} \middle| \frac{1}{2} n'_{e+}, \frac{1}{2} n'_{\nu_e} \right\rangle^2 = 1$.

Therefore, $\left\langle 1 \sqrt{2} \middle| \frac{1}{2} n'_{e+}, \frac{1}{2} n'_{\nu_e} \right\rangle_{rms} = 1$, within a phase factor.

Similarly, we can determine $\left\langle 1\, 0 \middle| \frac{1}{2} n'_f, \frac{1}{2} n'_{\bar{f}} \right\rangle^2$ from:

$$\left\langle 1\, 0 \middle| \frac{1}{2} n'_f, \frac{1}{2} n'_{\bar{f}} \right\rangle^2 = \int_{-\infty}^{\infty} \left\langle \frac{1}{2} x, \frac{1}{2} - x \middle| 1\, 0 \right\rangle^2 dx = \int_{-\infty}^{\infty} \frac{dx}{2\left[(\frac{1}{2}+x)!(\frac{1}{2}-x)!\right]^2} = 1 \quad (C5)$$

Integrating Eq. (C5) yields $\left\langle 1\, 0 \middle| \frac{1}{2} n'_f, \frac{1}{2} n'_{\bar{f}} \right\rangle^2 = 1$, and therefore $\left\langle 1\, 0 \middle| \frac{1}{2} n'_f, \frac{1}{2} n'_{\bar{f}} \right\rangle_{rms} = 1$.

Next, we determine the value of $\left\langle 0\, 0 \middle| \frac{1}{2} n'_f, \frac{1}{2} n'_{\bar{f}} \right\rangle$. We use the analytical expression for the square of the Clebsch-Gordan coefficient $\langle j_1 m_1, j_2 m_2 | j_3 m_3 \rangle$ with $j_3 = j_1 + j_2 - 1$ and $m_3 = m_1 + m_2$ [49]:

$$\langle j_1 m_1, j_2 m_2 | j_1 + j_2 - 1, m_3 \rangle^2 = \frac{4(j_2 m_1 - j_1 m_2)^2 (2j_1-1)!(2j_2-1)!(j_1+j_2+m_1+m_2-1)!(j_1+j_2-m_1-m_2-1)!}{(2j_1+2j_2)!(j_1+m_1)!(j_1-m_1)!(j_2+m_2)!(j_2-m_2)!} \quad (C6)$$

Choosing an arbitrary axis in the frame of the decaying spin-0 particle, the projections of the two product spins are given by $x$ and $-x$, respectively. Similar to above, $\left\langle 0\, 0 \middle| \frac{1}{2} n'_f, \frac{1}{2} n'_{\bar{f}} \right\rangle^2$ can be written by the following integral, where $\left\langle \frac{1}{2} x, \frac{1}{2} - x \middle| 0\, 0 \right\rangle^2$ is determined from Eq. (C6):

$$\left\langle 0\, 0 \middle| \frac{1}{2} n'_f, \frac{1}{2} n'_{\bar{f}} \right\rangle^2 = \int_{-\infty}^{\infty} \left\langle \frac{1}{2} x, \frac{1}{2} - x \middle| 0\, 0 \right\rangle^2 dx = \int_{-\infty}^{\infty} \frac{2x^2 \, dx}{\left[(\frac{1}{2}+x)!(\frac{1}{2}-x)!\right]^2} \quad (C7)$$

Equation (C7) can be brought into the form of Eq. (C4) by using $\Gamma(x+1) = x\Gamma(x)$, so that:

$$\int_{-\infty}^{\infty} \frac{2x^2 \, dx}{\left[(\frac{1}{2}+x)!(\frac{1}{2}-x)!\right]^2} = \int_{-\infty}^{\infty} \left[\frac{2}{\Gamma(\frac{3}{2}+x)\Gamma(\frac{1}{2}+x)\Gamma(\frac{3}{2}-x)\Gamma(\frac{1}{2}-x)} - \frac{(1/2)}{\Gamma(\frac{3}{2}+x)^2 \Gamma(\frac{3}{2}-x)^2}\right] dx = 1 \quad (C8)$$

Note that there is also an integral proportional to $x$ on the right-hand side (not shown), which is odd, and yields 0 for integration over the even domain. Evaluating Eq. (C8) yields $\left\langle 0\, 0 \middle| \frac{1}{2} n'_f, \frac{1}{2} n'_{\bar{f}} \right\rangle^2 = 1$, and therefore $\left\langle 0\, 0 \middle| \frac{1}{2} n'_f, \frac{1}{2} n'_{\bar{f}} \right\rangle_{rms} = 1$ (within a phase factor).

Finally, we determine the value of $\langle 0\,0|1\,n'_b, 1\,n'_{\bar{b}}\rangle$, using the analytical expression for the square of the Clebsch-Gordan coefficient $\langle j_1 m_1, j_2 m_2|j_3 m_3\rangle$ with $j_3 = j_1 + j_2 - 2$ and $m_3 = m_1 + m_2$ [49]:

$$\langle j_1 m_1, j_2 m_2|j_1 + j_2 - 2, m_3\rangle^2 = \frac{2j_1(2j_1-1)2j_2(2j_2-1)}{2(2j_1+2j_2-2)(2j_1+2j_2-1)}\left[\binom{2j_1}{j_1-m_1}\binom{2j_2}{j_2-m_2}\binom{2j_1+2j_2-4}{j_1+j_2-m_1-m_2-2}\right]^{-1}$$

$$\times \left[\binom{2j_1-2}{j_1-m_1}\binom{2j_2-2}{j_2+m_2} - 2\binom{2j_1-2}{j_1-m_1-1}\binom{2j_2-2}{j_2+m_2-1} + \binom{2j_1-2}{j_1-m_1-2}\binom{2j_2-2}{j_2+m_2-2}\right] \quad (C9)$$

Choosing an arbitrary axis in the frame of the decaying spin-0 particle, the projections of the two product spins are given by $x$ and $-x$, respectively. Similar to above, $\langle 0\,0|1\,n'_f, 1\,n'_{\bar{f}}\rangle^2$ can be written, using Eq. (C9), by the following integral:

$$\langle 0\,0|1\,n'_f, 1\,n'_{\bar{f}}\rangle^2 = \int_{-\infty}^{\infty}\langle 1\,x, 1\,-x|0\,0\rangle^2 dx = \int_{-\infty}^{\infty}\frac{(1-3x^2)^2 \sin^2(\pi x)\,dx}{3\pi^2 x^2 (1-x^2)^2} = 1 \quad (C10)$$

We plot the quasiprobability distributions of $\langle\frac{1}{2}x,\frac{1}{2}-x|1\,0\rangle^2$, $\langle\frac{1}{2}x,\frac{1}{2}\sqrt{2}-x|1\,\sqrt{2}\rangle^2$, and $\left\langle \frac{1}{2} x, \frac{1}{2} - x \middle| 0\ 0 \right\rangle^2$ in Fig. C1, using Eqs. (C1) and (C6), and expressed explicitly in Eqs. (C5), (C3), and (C7), respectively. The only physical observable of these distributions is their area, the total probability, which is 1. These are quasiprobability distributions because they can have regions of probability that are negative or greater than 1 [51,52,53]. For example, see Fig. C1b, where the peak around $x = \sqrt{2}/2$ has probability greater than 1, whereas the tails have negative probabilities (which exactly cancel excess probabilities near the peak). Notice also that panel (b) is the only one that gives a distribution that is symmetric about $x = \sqrt{2}/2$, (the rest are symmetric about $x = 0$). The reason for this is that the decaying particle has $n_3 = \sqrt{2}$ (the other three have $n_3 = 0$), and the two product particles have $n_1' = x$, $n_2' = \sqrt{2} - x$, $j_1 = j_2$, and $n_1' = n_2'$ for $x = \sqrt{2}/2$. Therefore, the distribution must be symmetric about $x = \sqrt{2}/2$. Note also that the classically allowed region is white, and the classically forbidden region is shaded grey.

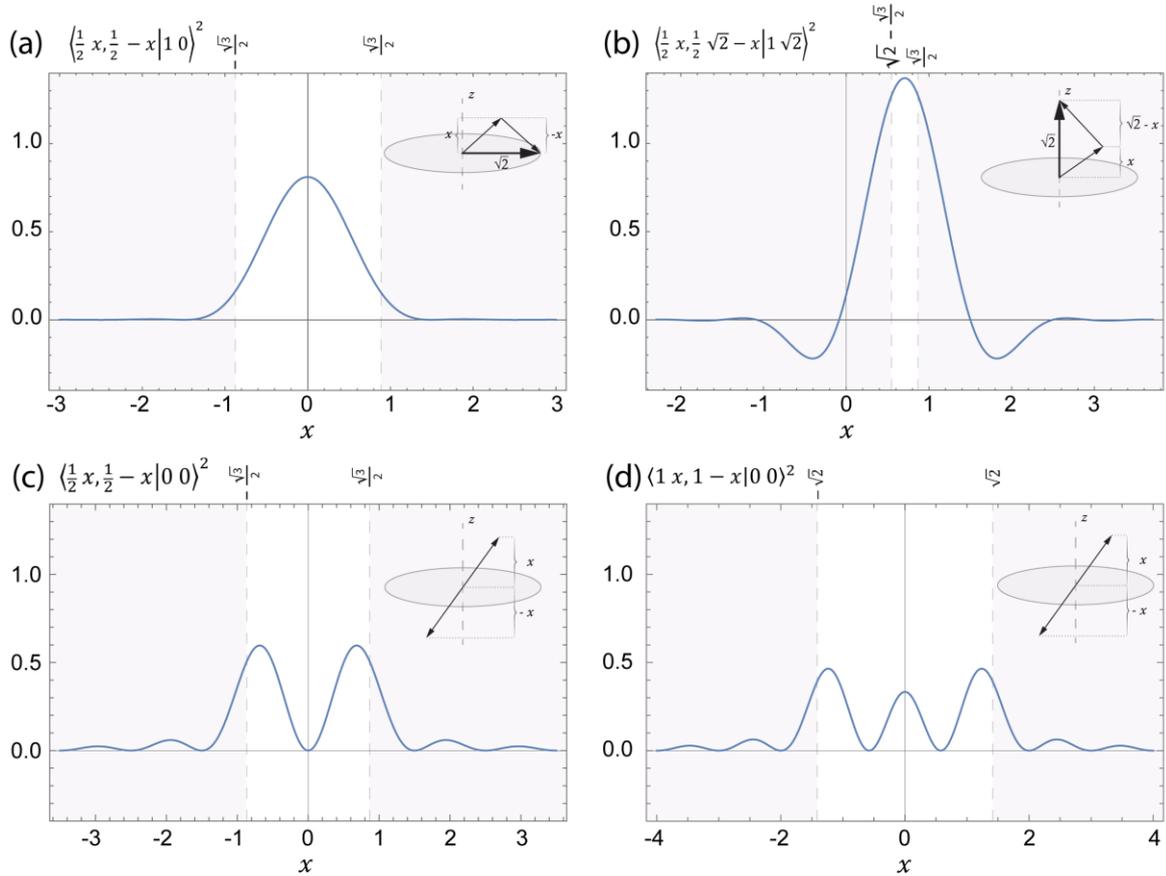

*Figure C1:* Quasiprobability distributions (a) $\langle\frac{1}{2}x,\frac{1}{2}-x|1\,0\rangle^2$, (b) $\langle\frac{1}{2}x,\frac{1}{2}\sqrt{2}-x|1\,\sqrt{2}\rangle^2$, (c) $\langle\frac{1}{2}x,\frac{1}{2}-x|1\,0\rangle^2$, and (d) $\langle\frac{1}{2}x,\frac{1}{2}-x|0\,0\rangle^2$ vs. $x$, which is the projection of one product spin on the internal-frame quantization axis of the decaying particle's spin. The other product spin has projection (a,c,d) $-x$, and (b) $\sqrt{2}-x$. The forbidden region, where $|\sqrt{2}-x| > |s_f|$ or $|x| > |s_f|$, is shaded grey. A vector-model pictorial description of the spin coupling is shown in the inset of each panel.

The projection $x$ is the projection of a produced spin $\boldsymbol{s}$ along the internal frame of the decaying particle, and is given by $x = |\boldsymbol{s}|cos\theta_s$, where $\theta_s$ is the angle between $\boldsymbol{s}$ and $z$. Note that $\theta_s$ is real in the classically allowed region $-|\boldsymbol{s}| < x < |\boldsymbol{s}|$, whereas $\theta_s$ is purely imaginary in the classically forbidden region $-|\boldsymbol{s}| > x > |\boldsymbol{s}|$ (similar to how the momentum of a particle is imagery in the classically forbidden region). Our expectation is that the probability will oscillate in the classically allowed region, and will decay exponentially in the classically forbidden region, as shown in Fig. C1. Notice that in the inset of each panel of Fig. C1 a vector-model pictorial description is given, to visualize the projections of each angular momentum along the $z$ axis. The decaying particle spin has a fixed projection along $z$ (as this is the decaying particle frame), whereas the two product particle spins have projections that range from $-\infty$ to $+\infty$.

# REFERENCES

[1] D. J. Griffiths, "Introduction to Quantum Mechanics," 2nd Edition, Pearson Prentice Hall, Upper Saddle River, New Jersey, 2005.

[2] S. E. Choi and R. B. Bernstein, "Theory of oriented symmetric-top molecule beams: Precession, degree of orientation, and photofragmentation of rotationally state-selected molecules", *J. Chem. Phys.* **85**, 150 (1986), DOI: 10.1063/1.451821

[3] R.D. Levine, R.B. Bernstein, "Rotational state dependence of the reactivity of oriented symmetric top molecules", *Chem. Phys. Lett.* **132**, 11 (1986). DOI: 10.1016/0009-2614(86)80685-6

[4] R. N. Zare, "Photofragment angular distributions from oriented symmetric-top precursor molecules", *Chem. Phys. Lett.* **156**, 1 (1989), DOI: 10.1016/0009-2614(89)87070-8

[5] Suketu R. Gandhi, Thomas J. Curtiss, Qi-Xun Xu, Seung E. Choi, Richard B. Bernstein, "Oriented molecule beams: pulsed, focused beams of methyl halides in pure JKM rotational states", *Chem. Phys. Lett.* **132**, 6 (1986), DOI: 10.1016/0009-2614(86)80684-4

[6] E. W. Kuipers, M. G. Tenner, A. W. Kleyn & S. Stolte, "Observation of steric effects in gas–surface scattering", *Nature* **334**, 420 (1988), DOI: 10.1038/334420a0

[7] T.P. Rakitzis, A.J. van den Brom, M.H.M. Janssen, "Directional dynamics in photodissociation of oriented molecules", *Science* **303**, 1852 (2004), DOI: 10.1126/science.1094186

[8] F. Wang, K. Liu, & T. P. Rakitzis, "Revealing the stereospecific chemistry of the reaction of Cl with aligned $CHD_3(v_1 = 1)$", *Nature Chem.* **4**, 636 (2012), DOI: 10.1038/nchem.1383

[9] T. P. Rakitzis, M. E. Koutrakis, G. E. Katsoprinakis, "The "Vector-Model" Wavefunction: spatial description and wavepacket formation of quantum-mechanical angular momenta", *Phys. Scr.* **99**, 075401 (2024). DOI: 10.1088/1402-4896/ad4ea1

[10] D. K. Sunko, "Angular momentum in two dimensions", *Eur. J. Phys.* **31**, L59–L64 (2010), DOI: 10.1088/0143-0807/31/3/L03

[11] D. Pandres and D. A. Jacobson, "Scalar Product for Harmonic Functions of the Group SU(2)", *J. Math. Phys.* **9**, 1401 (1968).

[12] S. Ferrara, M. Porrati, V. L. Telegdi, "*g=2 as the natural value of the tree-level gyromagnetic ratio of elementary particles*", *Phys. Rev. D* **46**, 3529 (1992), DOI: 10.1103/PhysRevD.46.3529

[13] B. R. Holstein, "*How Large is the "Natural" Magnetic Moment?*", *Am. J. Phys.* **74**, 1104–1111 (2006), DOI: 10.1119/1.2345655

[14] S. Weinberg, "*The Quantum Theory of Fields, Volume 1: Foundations*" Cambridge University Press, 2005, ISBN: 9781139644167.

[15] M. Pavšič, "Manifestly covariant canonical quantization of the scalar field and particle localization", Mod. Phys. Lett. A **33**, 1850114 (2018). DOI: 10.1142/S0217732318501146

[16] M. Pavšič, "Localized States in Quantum Field Theory", arXiv:1705.02774 (2018).

[17] M. Pavšič, "A new perspective on quantum field theory revealing possible existence of another kind of fermions forming dark matter", Int. J. Geom. Methods Mod. Phys. **19**, 2250184 (2022). DOI: 10.1142/S0219887822501845.

[18] R. N. Zare, "*Angular Momentum*", Wiley, New York, 1988, ISBN: 978-0-471-85892-8

[19] D. Pandres, "Schrodinger Basis for Spinor Representations of the Three-Dimensional Rotation Group", *J. Math. Phys.* **6**, 1098 (1965).

[20] M. Pavsic, "Rigid Particle and its Spin Revisited" Found. Phys. **37**, 40 (2007), DOI: 10.1007/s10701-006-9094-4

[21] J. J. Sakurai, "*Advanced Quantum Mechanics*", Addison-Wesley, 1967, ISBN: 9780201067101

[22] A. R. Edmonds, "*Angular Momentum in Quantum Mechanics*", Princeton University Press, 1974, ISBN: 9780691025896

[23] J.-M. Levy-Leblond, "*Nonrelativistic particles and wave equations*", Commun. Math. Phys., 6, 286-311 (1967), DOI: 10.1007/BF01646020

[24] J.D. Bjorken and S.D. Drell, "Relativistic Quantum Mechanics", McGraw-Hill, New York (1964)

[25] A. Proca, "Sur les Equations Fondamentales des Particules Elementaires," Comp. Ren. Acad. Sci. Paris **202**, 1366 (1936).

[26] W. Greiner, "Relativistic quantum mechanics", Springer-Verlag, Berlin, 2000, ISBN: 3-540-67457-8

[27] F.J. Belinfante, "Intrinsic Magnetic Moment of Elementary Particles of Spin 3/2," Phys. Rev. **92**, 997 (1953), DOI: 10.1103/PhysRev.92.997

[28] Particle Data Group, S. Eidelman et al., "Review of Particle Physics", Phys. Lett. **B592**, 1 (2004), DOI: 10.1016/j.physletb.2004.06.001

[29] I.J.R. Aitchison and A.J.G. Hey, "Gauge Theories in Particle Physics", Adam Hilger, Philadelphia (1989), ISBN: 0-85274-329-7

[30] L. D. Landau, "The moment of a 2-photon system", Dokl. Akad. Nauk SSSR. **60**, 207 (1948).

[31] C. N. Yang, "Selection Rules for the Dematerialization of a Particle into Two Photons", Phys. Rev. **77**, 242 (1950), DOI: 10.1103/PhysRev.77.242

[32] T.R. Govindarajan, Rakesh Tibrewala, "Generalized Landau Yang Theorem", arXiv:2410.04786 (2024).

[33] R. Foot, "Mirror Matter-Type Dark Matter", Int. J. Mod. Phys. D 13, 2161 (2004), DOI: 10.1142/S0218271804006449

[34] Ade et al. (Planck Collaboration), "Planck 2013 results. XVI. Cosmological parameters",A&A **571**, A16 (2014), DOI: 10.1051/0004-6361/201321591

[35] P. Ciarcelluti, Q. Wallemacq, "Is dark matter made of mirror matter? Evidence from cosmological data", *Phys. Lett. B* **729**, 62 (2014), DOI: 10.1016/j.physletb.2013.12.057

[36] J. S. Bullock and M. Boylan-Kolchin, , Ann. Rev. Astron. Astrophys., **55**, 343 (2017), DOI: 10.1146/annurev-astro-091916-055313

[37] F. Governato et al., "Bulgeless dwarf galaxies and dark matter cores from supernova-driven outflows", *Nature* **463**, 203 (2010), DOI: 10.1038/nature08640

[38] N. C. Relatores et al., "The Dark Matter Distributions in Low-mass Disk Galaxies. II. The Inner Density Profiles", ApJ **887**, 94 (2019), DOI: 10.3847/1538-4357/ab5305

[39] Ethan O. Nadler et al., "A Self-interacting Dark Matter Solution to the Extreme Diversity of Low-mass Halo Properties", ApJL **958**, L39 (2023), DOI: 10.3847/2041-8213/ad0e09

[40] J. Sánchez Almeida, I. Trujillo, and A. R. Plastino, *ApJL* **973**, L15 (2024). DOI: 10.3847/2041-8213/ad66bc

[41] A. Tumasyan et al. (CMS Collaboration) *Phys. Rev. D* **105**, 092007 (2022), DOI: 10.1103/PhysRevD.105.092007

[42] Y. Koide, "A New Formula For The Cabibbo Angle And Composite Quarks And Leptons," *Phys. Rev. Lett.* **47**, 1241 (1981), DOI: 10.1103/PhysRevLett.47.1241

[43] R. Foot, arXiv:hep-ph/9402242 (1994).

[44] G.H. Gao, N. Li, "Explorations of two empirical formulas for fermion masses", *Eur. Phys. J. C* **76**, 140 (2016), DOI: 10.1140/epjc/s10052-016-3990-3

[45] N. Li, B.-Q. Ma, "Estimate of neutrino masses from Koide's relation", *Phys. Lett. B* **609**, 309 (2005), DOI: 10.1016/j.physletb.2005.01.066

[46] A. Kartavtsev, "A remark on the Koide relation for quarks", arXiv:1111.0480 (2011).

[47] H. Ruegg, "A simple derivation of the quantum Clebsch-Gordan coefficients for SU(2)$_q$", *J. Math. Phys.* **31**, 1085 (1990). DOI: 10.1063/1.528787

[48] M. J. Caola, "A simple short derivation of the Clebsch-Gordan coefficients", *Lat. Am. J. Phys. Educ.* **4**, 84 (2010). ISSN 1870-9095

[49] D. A. Varshalovich, A. N. Moskalev, and V. K. Khersonskii, "*Quantum Theory of Angular Momentum*", World Scientific, Singapore, 1988, ISBN: 978-9971-5-0107-5

[50] I. S. Gradshteyn and I. M. Ryzhik, "*Table of Integrals, Series, and Products*", Academic Press, Amsterdam, 2007, ISBN-13: 978-0123736376

[51] M. Hillery, R. F. O'Connell, M. O. Scully, E. P. Wigner, "Distribution functions in physics: Fundamentals", *Phys. Rep.* **106**, 121–167 (1984). DOI: 10.1016/0370-1573(84)90160-1

[52] Richard P. Feynman, "Negative probabilities," in Quantum Implications: Essays in Honor of David Bohm, ed. F. D. Peat and B. Hiley, Routledge & Kegan Paul (1987) 235–248.

[53] Nick Polson, Vadim Sokolov, "Negative Probability", arXiv:2405.03043 (2024).